\documentclass{aa}

\usepackage{graphicx}
\usepackage{txfonts}
\usepackage{color}

\begin{document}

\title{Merging galaxy clusters in IllustrisTNG}

\author{Ewa L. {\L}okas
}

\institute{Nicolaus Copernicus Astronomical Center, Polish Academy of Sciences,
Bartycka 18, 00-716 Warsaw, Poland\\
\email{lokas@camk.edu.pl}}

\titlerunning{Merging galaxy clusters in IllustrisTNG}
\authorrunning{E. L. {\L}okas}


\abstract{
Mergers between galaxy clusters are
an important stage in the formation of the large-scale structure of the Universe. Some of the mergers show a
spectacular bow shock that formed as a result of recent passage of a smaller cluster through a bigger one, the classic
example of this being the so-called bullet cluster. In this paper, I describe ten examples of interacting clusters
identified among 200 of the most massive objects, with total masses above $1.4 \times 10^{14}$ M$_{\odot}$,
from the IllustrisTNG300 simulation by searching for prominent bow shocks in their temperature maps. Despite different
mass ratios of the two merging clusters, the events are remarkably similar in many respects. In all cases, the
companion cluster passed close to the main one only once, between 0.9 and 0.3 Gyr ago, with the pericenter distance of
100-530 kpc and a velocity of up to 3400 km s$^{-1}$. The subcluster, typically an order of magnitude smaller in mass
than the main cluster before the interaction, loses most of its dark matter and gas in the process. The displacement
between the collisionless part of the remnant and the bow shock is such that the remnant typically lags behind the
shock or coincides with it, with a single exception of the merger occurring with the largest velocity. Usually about
1\% of the gas cells in the merging clusters are shocked, and the median Mach numbers of these gas cells are around
two. Due to the relatively small size of the simulation box, no close analog of the bullet cluster was found, but I
identified one case that is similar in terms of mass, velocity, and displacement. The presented cases bear more
resemblance to less extreme observed interacting clusters such as A520 and Coma.}

\keywords{galaxies: evolution -- galaxies: interactions -- galaxies: clusters: general -- galaxies: clusters:
intracluster medium -- X-rays: galaxies: clusters -- dark matter}

\maketitle

\section{Introduction}

The bullet cluster 1E 0657-56 was the first clear example of a merging galaxy cluster showing a bow shock associated
with a subcluster that recently passed through a bigger cluster \citep{Markevitch2002}. The shock front was
propagating in front of a bullet-like gas cloud, which gave the structure its name. The cluster then attracted a lot of
attention, as it was claimed to provide good evidence for the presence of dark matter in galaxy clusters
\citep{Clowe2004, Clowe2006}. The argument was based on the reconstruction of the dominant mass distribution from weak
gravitational lensing that turned out to coincide with the concentration of galaxies rather than the gas. Since the
gas constitutes most of the baryons, this means that most of the mass is contributed by a collisionless
component in the form of dark matter distributed in a manner similar to the galaxies.

Discoveries of similar structures soon followed in other merging clusters, including A754
\citep{Markevitch2003}, A520 \citep{Markevitch2005}, A2146 \citep{Rodriguez2011}, A4067 \citep{Chon2015}, ZwCl
0008.8+5215 \citep{Golovich2017}, and A2399 \citep{Lourenco2020}. Still, the exact properties of the bullet cluster make
it a very rare occurrence among structures formed in the $\Lambda$CDM universe. In particular, \citet{Markevitch2004}
estimated the relative velocity of the two components of the bullet cluster to be on the order of 4500 km s$^{-1}$, and
such high velocities were initially claimed to be difficult to reconcile with the $\Lambda$CDM model \citep{Hayashi2006}.
The controversy now seems to be resolved, and the bullet cluster is no longer perceived as a challenge to the
generally accepted cosmological model \citep{Thompson2015}. The statistics of such rare objects have been explored using
cosmological simulations of the formation of the large-scale structure of the Universe, and it has been determined that the probability of
finding a massive halo pair with such a high relative velocity is indeed very low and the required volume should be at
least on the order of a few Gpc$^3$ in order to properly reproduce the tail of the pairwise velocity distribution
\citep{Bouillot2015}.

The detailed properties of the bullet cluster were nonetheless successfully reproduced in non-cosmological controlled
simulations of mergers between galaxy clusters of different masses \citep{Springel2007, Mastropietro2008, Lage2014}. In
such simulations, the bigger cluster typically had a mass on the order of $10^{15}$ M$_\odot$, while the bullet was
assumed to be a few times less massive. The adopted relative velocities ranged between 2000 and 4000 km s$^{-1}$, and
the initial separation of the clusters was on the order of a few megaparsecs. The simulations typically
employed simple ways to model the gas physics, such as those implemented in the Gadget-2 code combined with smoothed
particle hydrodynamics used by \citet{Springel2007}, which turned out to be sufficient to reproduce the main features
of the merging clusters. Some improvements, such as the effective viscosity of the gas and its thermal conductivity,
might be introduced, but there are many other parameters that could also be varied, including the orbital angular
momentum of the interaction or the baryon fraction of clusters.

Only recently has studying merging galaxy clusters with sufficient resolution and in a cosmological context become
possible. In this work,
I use the IllustrisTNG simulations of galaxy formation \citep{Springel2018, Marinacci2018, Naiman2018, Nelson2018,
Pillepich2018} that follow the evolution of dark matter and baryons in boxes of different sizes by solving for gravity
and magnetohydrodynamics with the moving mesh code \textsc{arepo} \citep{Springel2010}. The code attempts to
eliminate the weaknesses identified in commonly applied techniques such as smoothed particle hydrodynamics or adaptive
mesh refinement. The simulations include prescriptions for star formation, stellar evolution, chemical enrichment,
primordial and metal-line cooling of the gas, and stellar feedback with galactic outflows as well as black hole
evolution and feedback.

The evolution of galaxy clusters has already been addressed by a number of studies using Illustris and IllustrisTNG
simulations. For example, \citet{Pillepich2018} described the stellar content of clusters,
\citet{Gupta2018} studied the chemical pre-processing of cluster galaxies, and \citet{Barnes2018} investigated the cool-core
galaxy clusters. Other studies focused on the evolution of galaxies in clusters, with \citet{Yun2019} discussing
gas-stripping phenomena and jellyfish galaxies, \citet{Sales2020} the formation of ultra-diffuse galaxies in clusters,
\citet{Joshi2020} the fate of disk galaxies in clusters, \citet{Lokas2020} the tidal evolution of galaxies in a massive
cluster, and \citet{Dacunha2022} the signatures of the evolutionary history in the observable properties of galaxies.

\begin{figure}
\centering
\includegraphics[width=9cm]{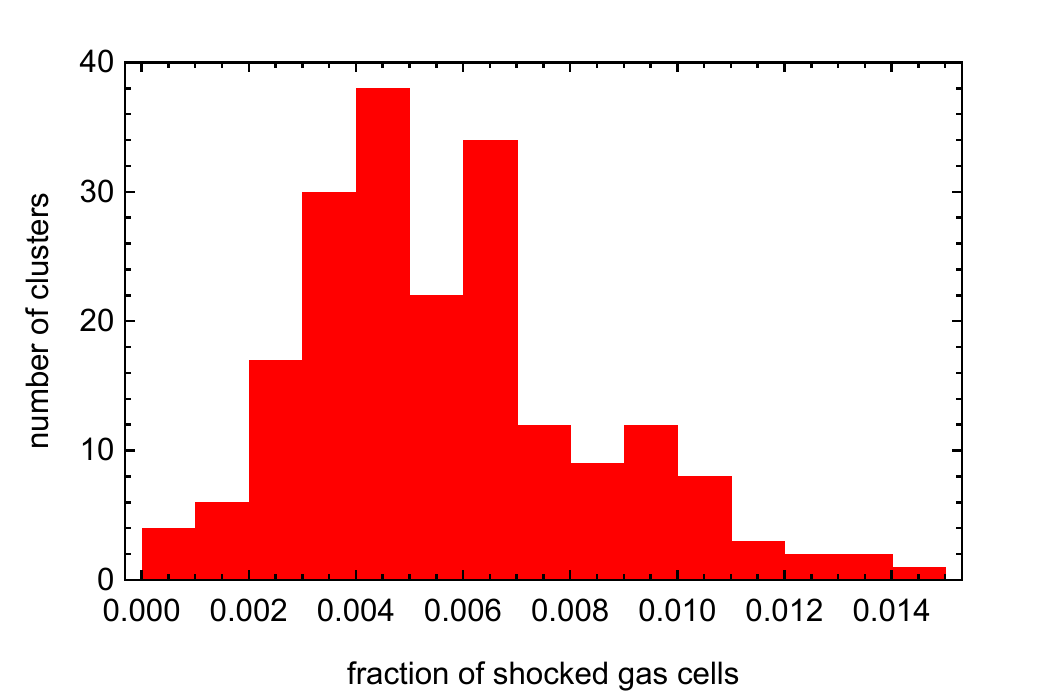}
\caption{Distribution of the fraction of shocked gas cells in 200 most massive clusters of IllustrisTNG300.}
\label{histogramshocked}
\end{figure}

\begin{figure*}
\centering
\includegraphics[width=6.25cm]{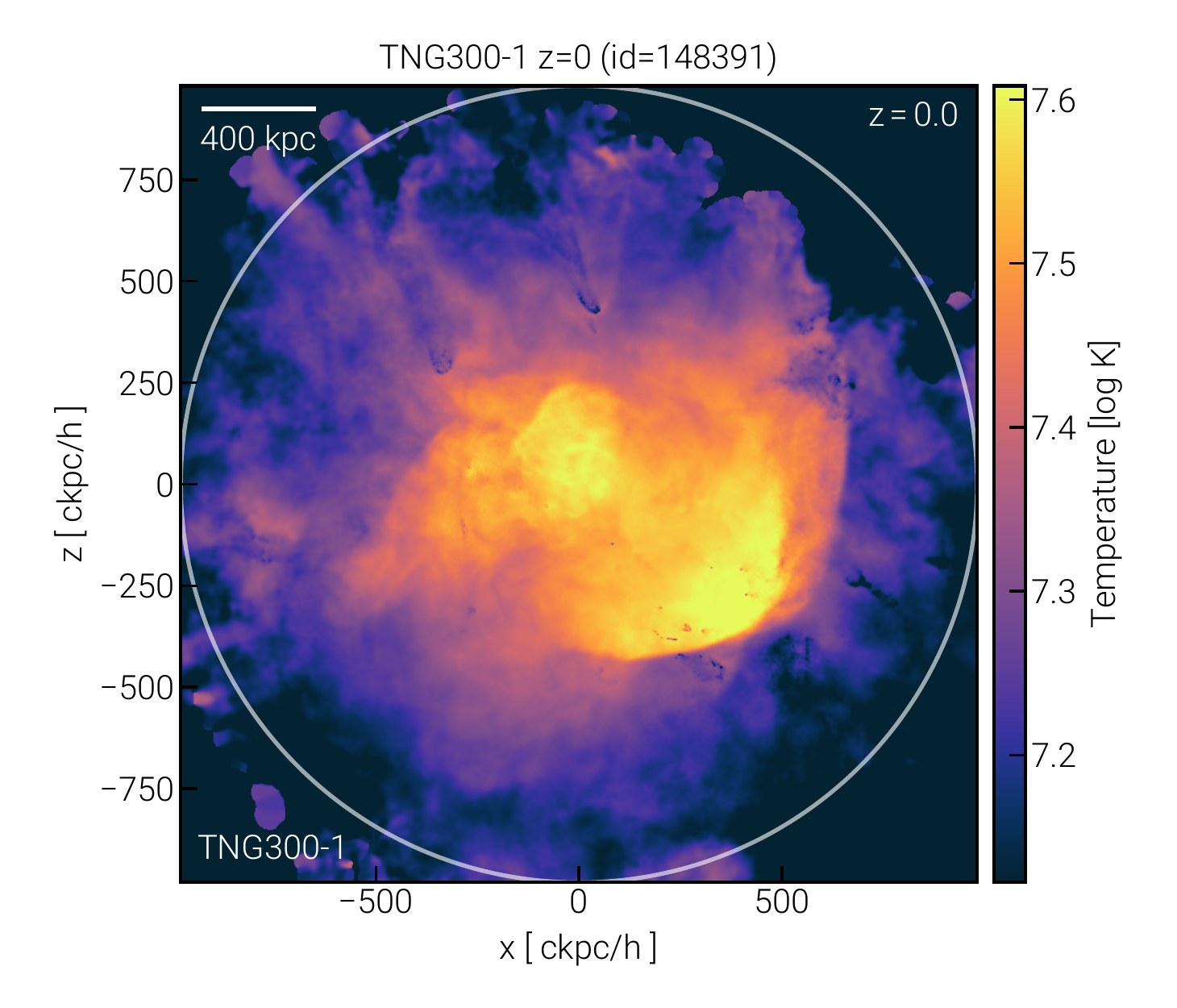}
\hspace{-0.4cm}
\includegraphics[width=6.25cm]{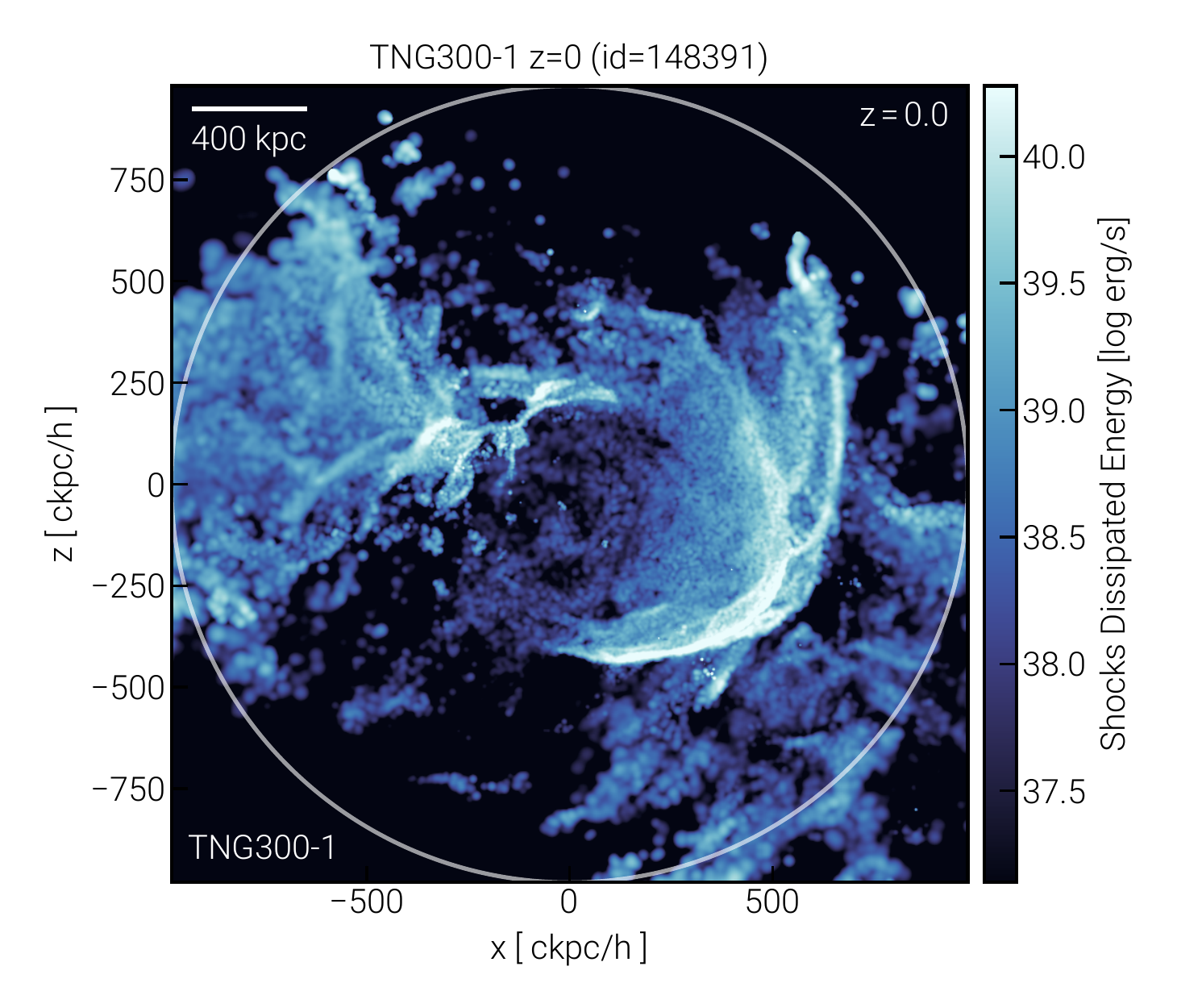}
\hspace{-0.4cm}
\includegraphics[width=6.25cm]{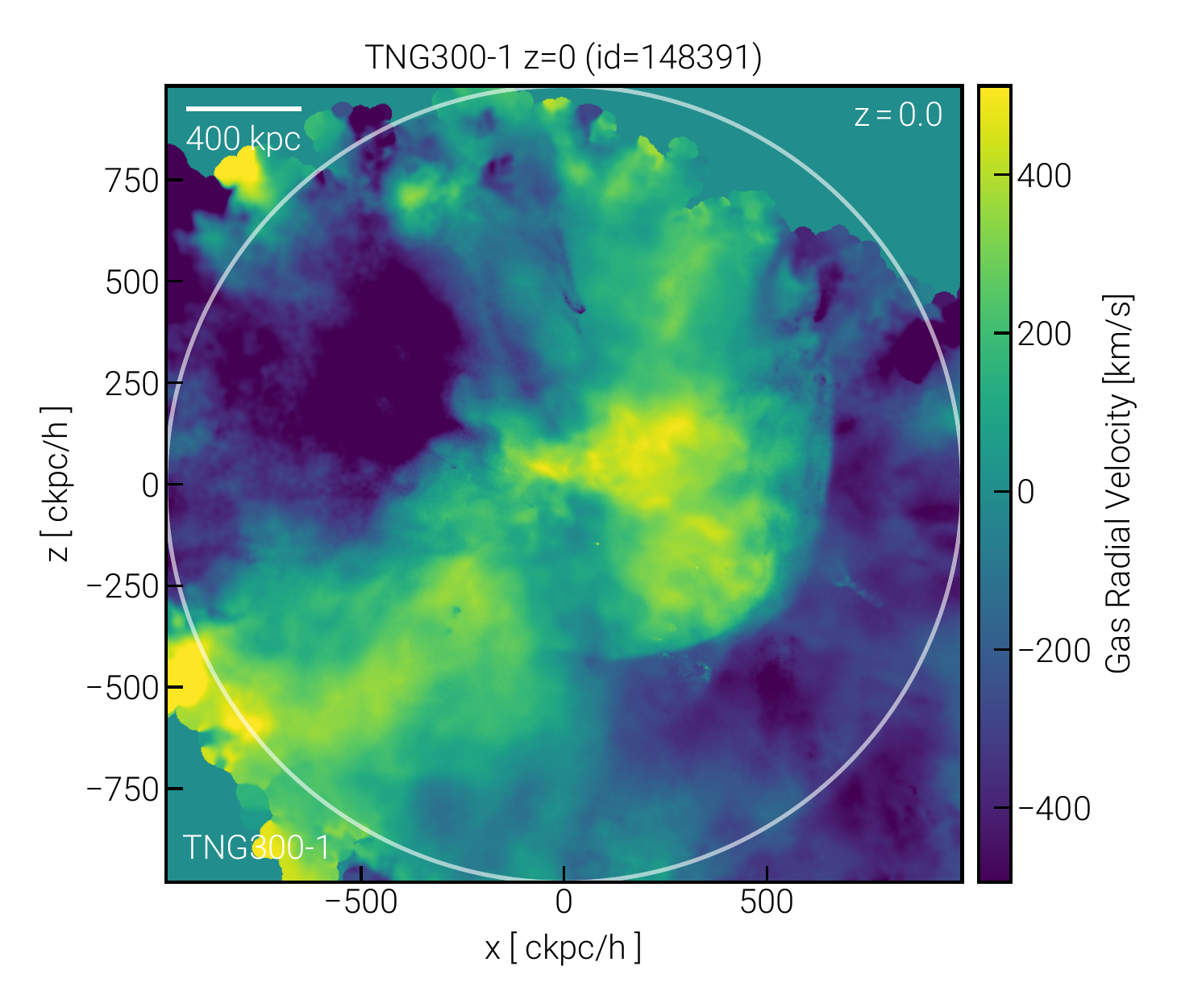}
\\
\includegraphics[width=6.25cm]{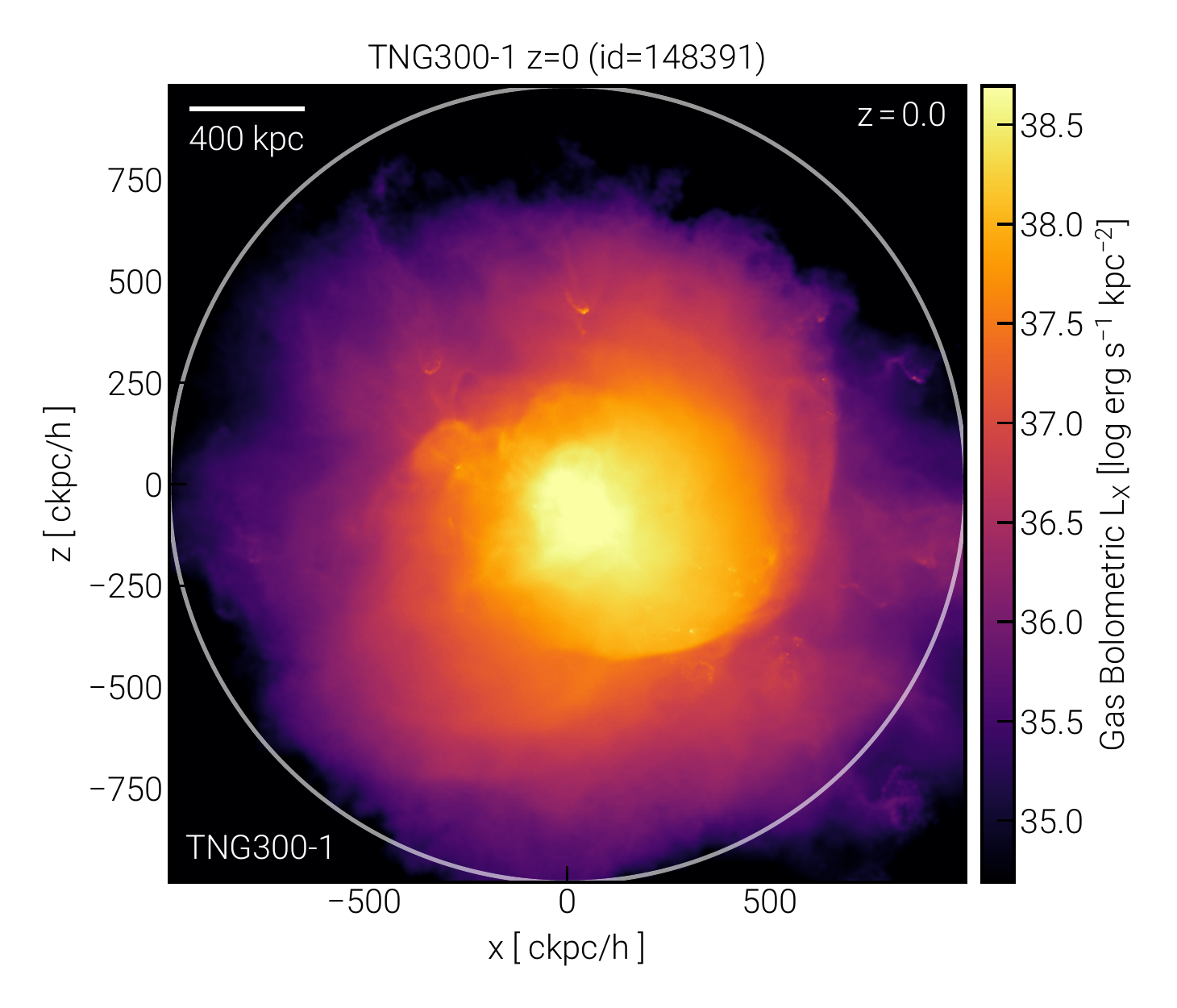}
\hspace{-0.4cm}
\includegraphics[width=6.25cm]{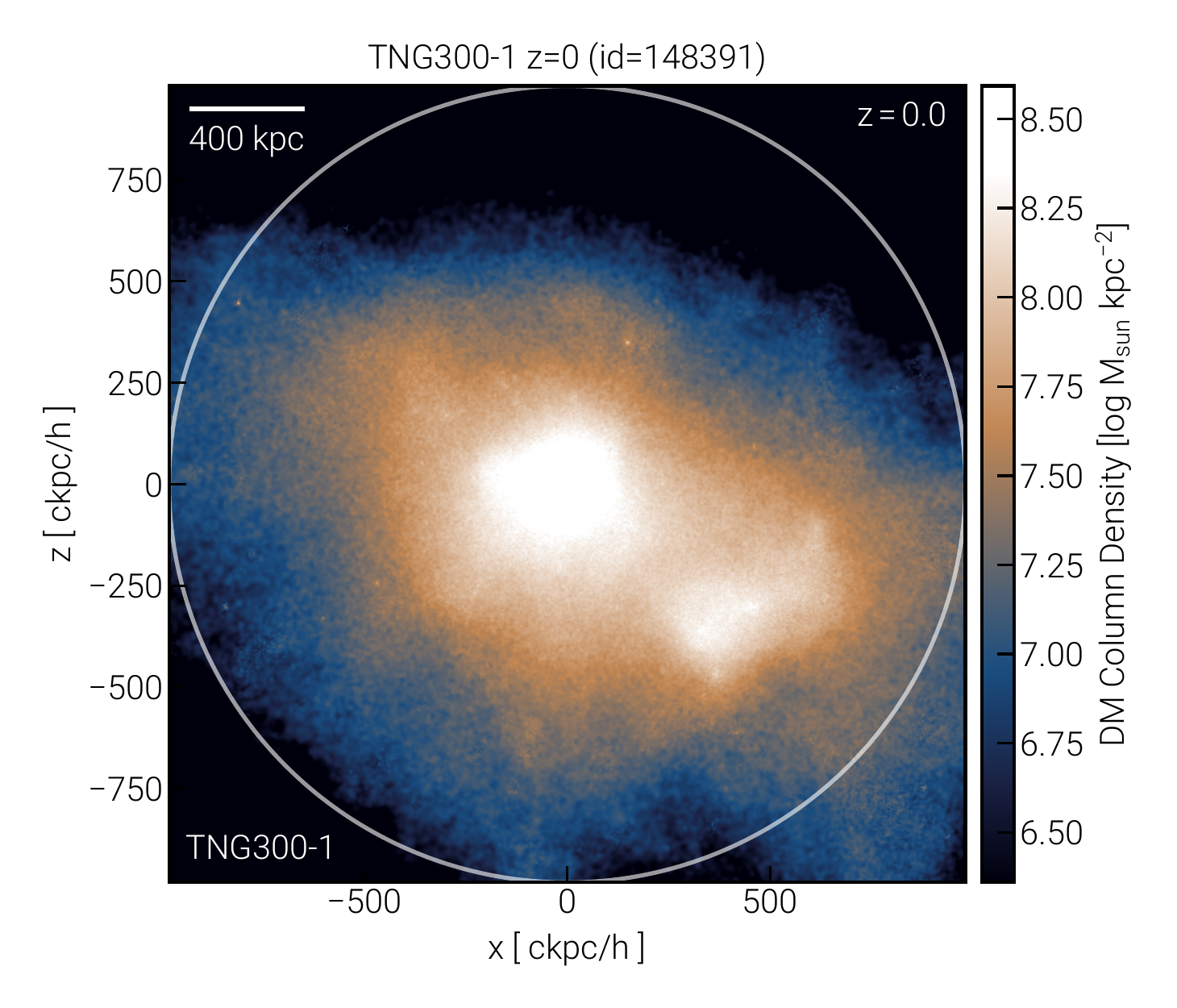}
\hspace{-0.4cm}
\includegraphics[width=6.25cm]{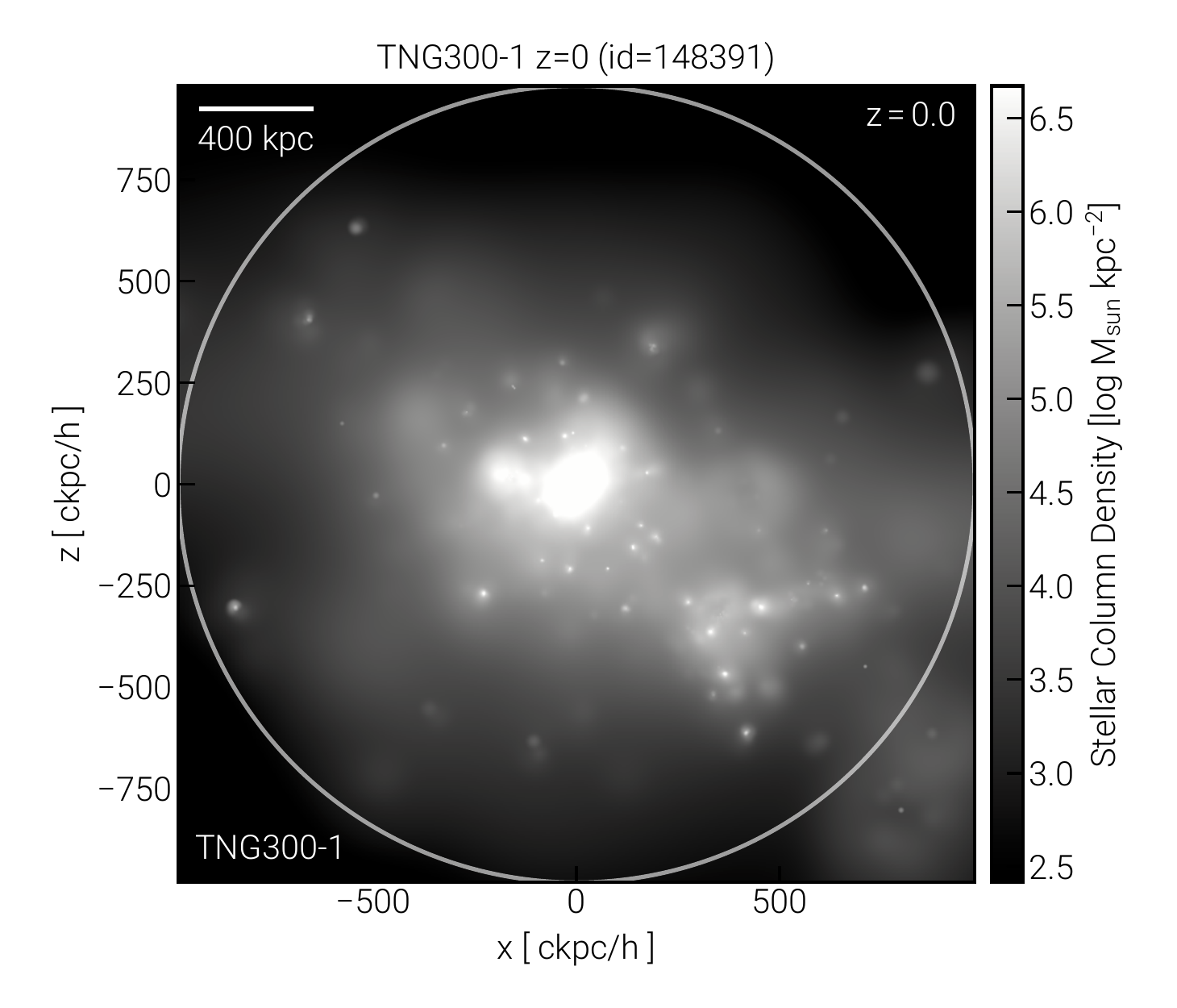}
\caption{Maps of different properties of the interacting clusters 148391-148395 at $z=0$. The subsequent panels, from
the upper left to the lower right, show the gas temperature, the energy dissipated in shocks, the gas radial velocity,
the gas X-ray luminosity, the dark matter density, and the stellar density. The white circles indicate the virial
radius of the bigger cluster.}
\label{images}
\end{figure*}

Cluster mergers have not yet been studied in detail in cosmological simulations, except for a few cases
selected from Illustris-NR-2 by \citet{Schaal2015}. In this paper, I report the results of the first systematic
study aimed at identifying convincing examples of merging clusters in IllustrisTNG and investigate their properties.
For this purpose, the most useful run is the TNG300 simulation in the box of 300 Mpc on a side, which contains
a sufficient number of massive clusters to search for merging objects in the final output. Interestingly, the
resolution of this run, in terms of the particle masses and softening scales, is comparable or better than the one
available in controlled simulations performed a decade earlier. Namely, the TNG300 run employs softenings of $0.4-1.5$
kpc and particle masses of $(1-7) \times 10^7$ M$_\odot$, which translates into at least a few million particles per
cluster. Although the 300 Mpc box size is still much smaller than what would be required to find a close analog of the
bullet cluster, it turns out to be sufficient for identifying about ten convincing examples of merging objects among
the most massive clusters. Section~2 presents the criteria and procedures used to identify the merging clusters in
IllustrisTNG. In Section~3, I describe in detail a typical example of these clusters, and in Section~4, I provide six
more cases, including one that is the most similar to the bullet cluster. Section~5 presents three special cases of
equal-mass mergers, and the discussion follows in the last section.

\section{Selection of merging clusters}

For this study, I used the highest resolution run TNG300-1 of the set with the 300 Mpc box, one of the publicly
available simulations of the IllustrisTNG, described by \citet{Nelson2019}. I searched for merging clusters among the
200 most massive subhalos with total mass $M_{\rm tot} > 1.4 \times 10^{14}$ M$_{\odot}$ identified in the last
simulation snapshot corresponding to the present time $t_0=13.8$ Gyr ($z=0$). The search was performed through visual
inspection of the temperature maps of the clusters along different lines of sight and by selecting the objects
showing a region of enhanced temperature in a shape of a bow shock within the cluster virial radius.

Since the shocks are associated with the presence of gas cells with Mach numbers $\mathcal{M} > 1$, one might expect
that a more systematic (and automatic) way to identify merging clusters would be to find those with the largest
fraction of shocked gas cells, $f_{\rm sh}$. The distribution of these fractions for the 200 most massive clusters in
the TNG300 box is shown in Fig.~\ref{histogramshocked}. It turns out there is a significant number of objects with
$f_{\rm sh}$ on the order of 0.01 or higher that do not make it into the sample of merging clusters. In contrast, the
selected merging clusters show a variety of values for $f_{\rm sh}$ in the range of 0.005-0.012, which is still
comparable or larger than the median of the distribution equal to 0.005.

For each of the identified cases of merging clusters, I looked for the most massive neighbors that interacted with the
main cluster in the last few Gyr and studied their orbits to identify the object most likely to have produced the bow
shock. I then followed the mass evolution of the subclusters by extracting the mass content of each subhalo in
subsequent simulation snapshots, as determined using the Subfind algorithm \citep{Springel2001}, which ensures that the
identified structures are self-bound. During the interaction, the subcluster underwent heavy tidal stripping
and lost most of its gas and dark matter. As a final identity check, I verified if the majority of the gas particles
that belonged to the subcluster at the time it reached its maximum mass were included among the gas particles of the
main cluster in the final simulation output. Although there are often a few massive galaxies interacting with the main
cluster around the same time, it was quite straightforward to identify the most massive galaxy actually responsible for
the creation of the bow shock using this procedure.

It is worth noting that the mergers cannot be identified by studying the merger trees of the biggest clusters provided
in the IllustrisTNG data. This is because, as demonstrated below, at the time the bow shocks are best visible, the
subclusters have not yet fully merged with the bigger objects. Although the smaller clusters lost
most of their dark matter and gas at the pericenter passage, they retained the stellar cores and identity, and
therefore are included in the subhalo catalogs as separate objects.

\section{A flagship example of merging clusters}

The most convincing example of merging clusters where a subcluster has recently passed through a bigger structure and
has formed a clear bow shock is shown in Fig.~\ref{images}, in the projection where the bow shock is best visible. The
cluster and the subcluster are identified by the identification number of their most massive subhalo. In the subhalo
catalog of the simulation at $t_0$, the bigger cluster has the identification number id = 148391, and the smaller one
has id = 148395. The images in all panels are centered on the bigger subhalo. The upper-left panel of
Fig.~\ref{images} shows the temperature map of the interacting structures with the bow shock visible as the
umbrella-like shape in the lower-right part of the image, with the temperature reaching $4 \times 10^7$ K.

The presence of the shock is confirmed in the upper-middle panel of the figure, which shows the energy dissipated in
shocks, that is, the amount of kinetic energy transformed into thermal energy, calculated using the on-the-fly cosmic
shock finder coupled to the simulation code \citep{Schaal2015, Schaal2016}. The motion of the gas is clear from
the upper-right panel presenting its
radial velocity. The structure of the shock is also visible in the lower-left panel, which is a map of the
X-ray bolometric luminosity of the gas, a quantity directly accessible in observations. It has the form of a sudden
drop in the luminosity at the region corresponding to the border of the bow shock in the other panels.

To complete the picture, the lower-middle and lower-right panels show, respectively, the projected distributions of
dark matter and stars in the merging clusters. In both images, one can see the substructure in the lower right
corresponding to the group of galaxies that has passed through the main cluster. It is worth noting that although I
identify the subcluster by its most massive subhalo, it corresponds only to the most massive galaxy of the group, and
there are always other galaxies that accompany the most massive galaxy in the passage through the bigger
cluster.

\begin{figure*}
\centering
\includegraphics[width=6.25cm]{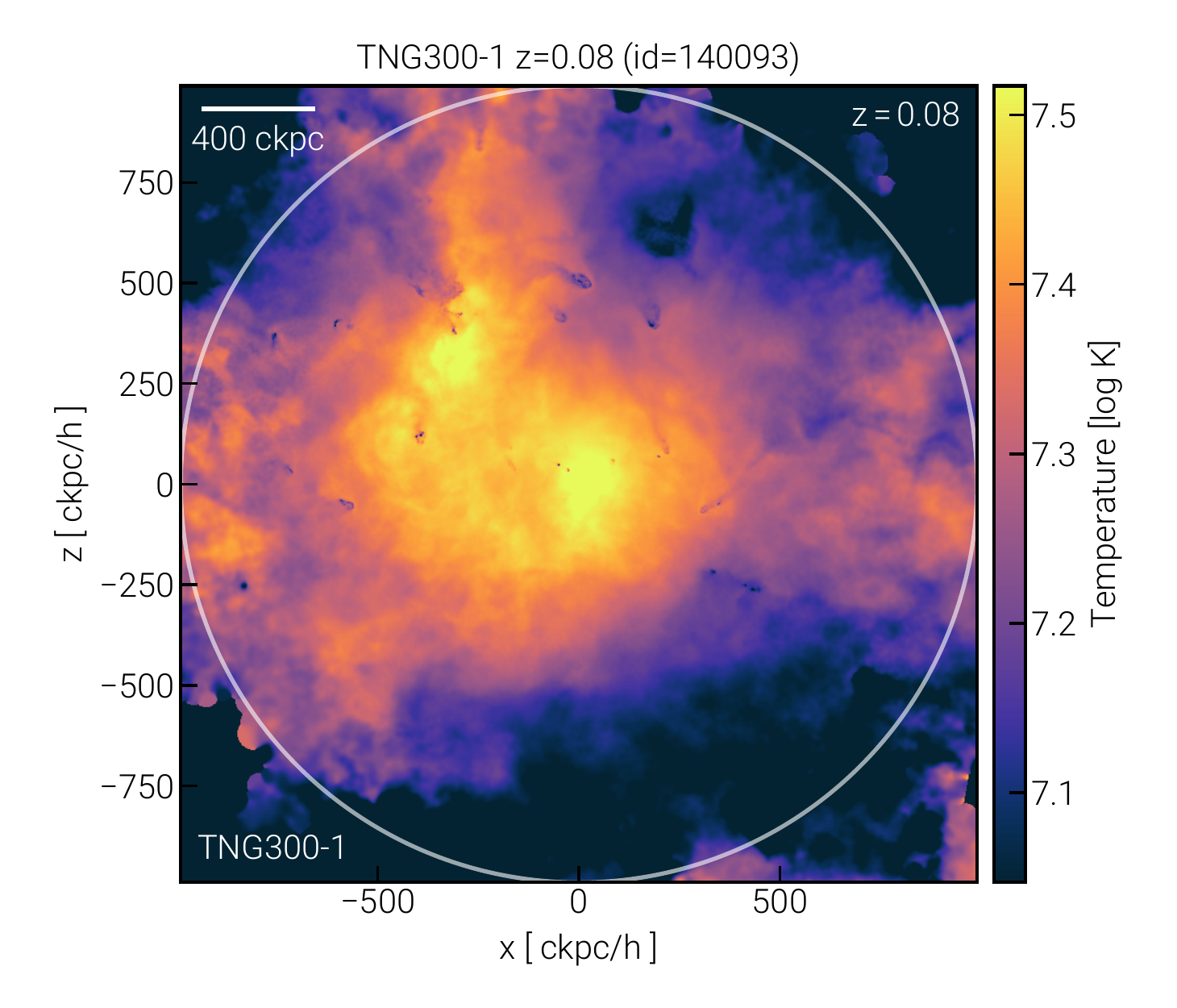}
\hspace{-0.4cm}
\includegraphics[width=6.25cm]{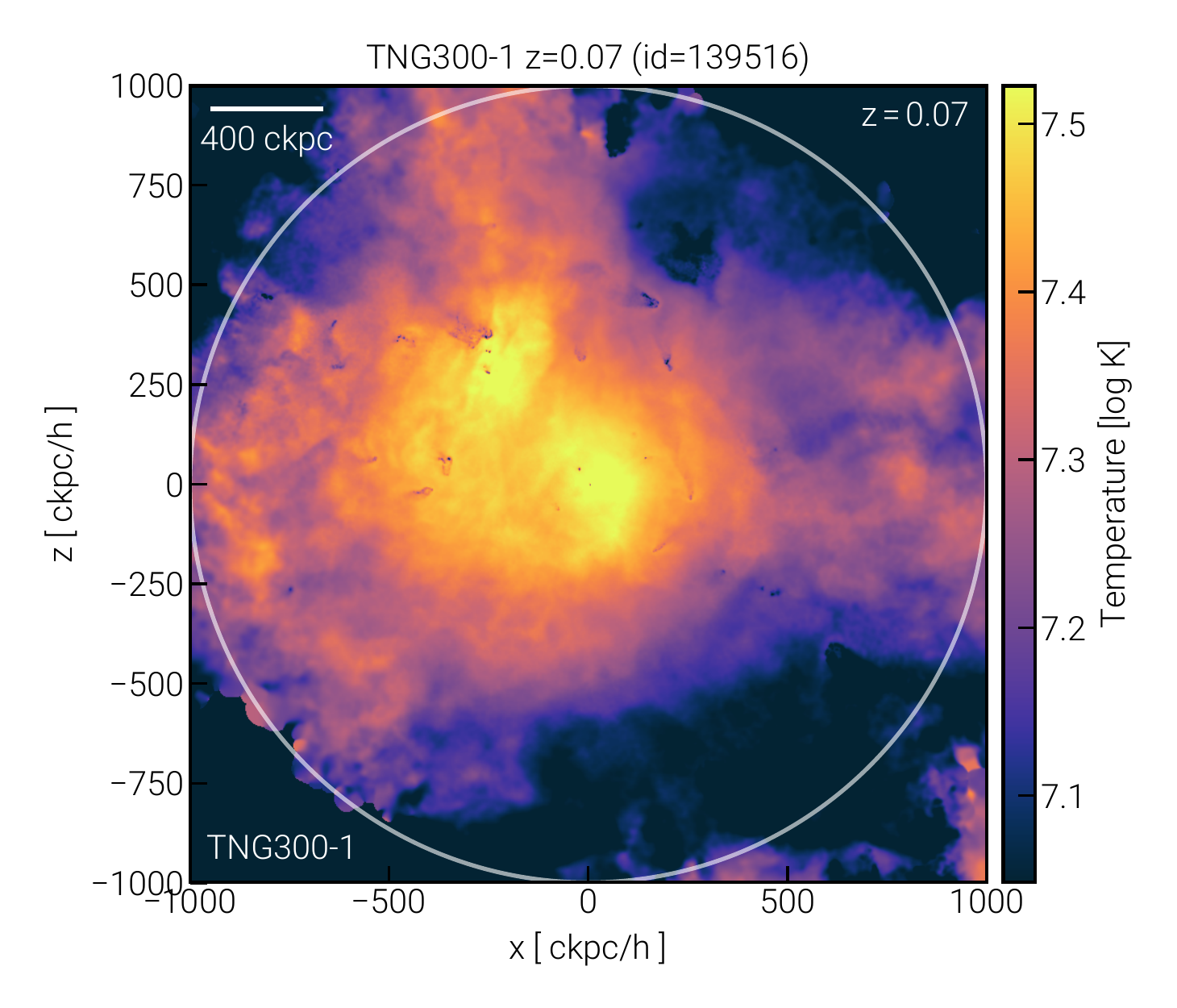}
\hspace{-0.4cm}
\includegraphics[width=6.25cm]{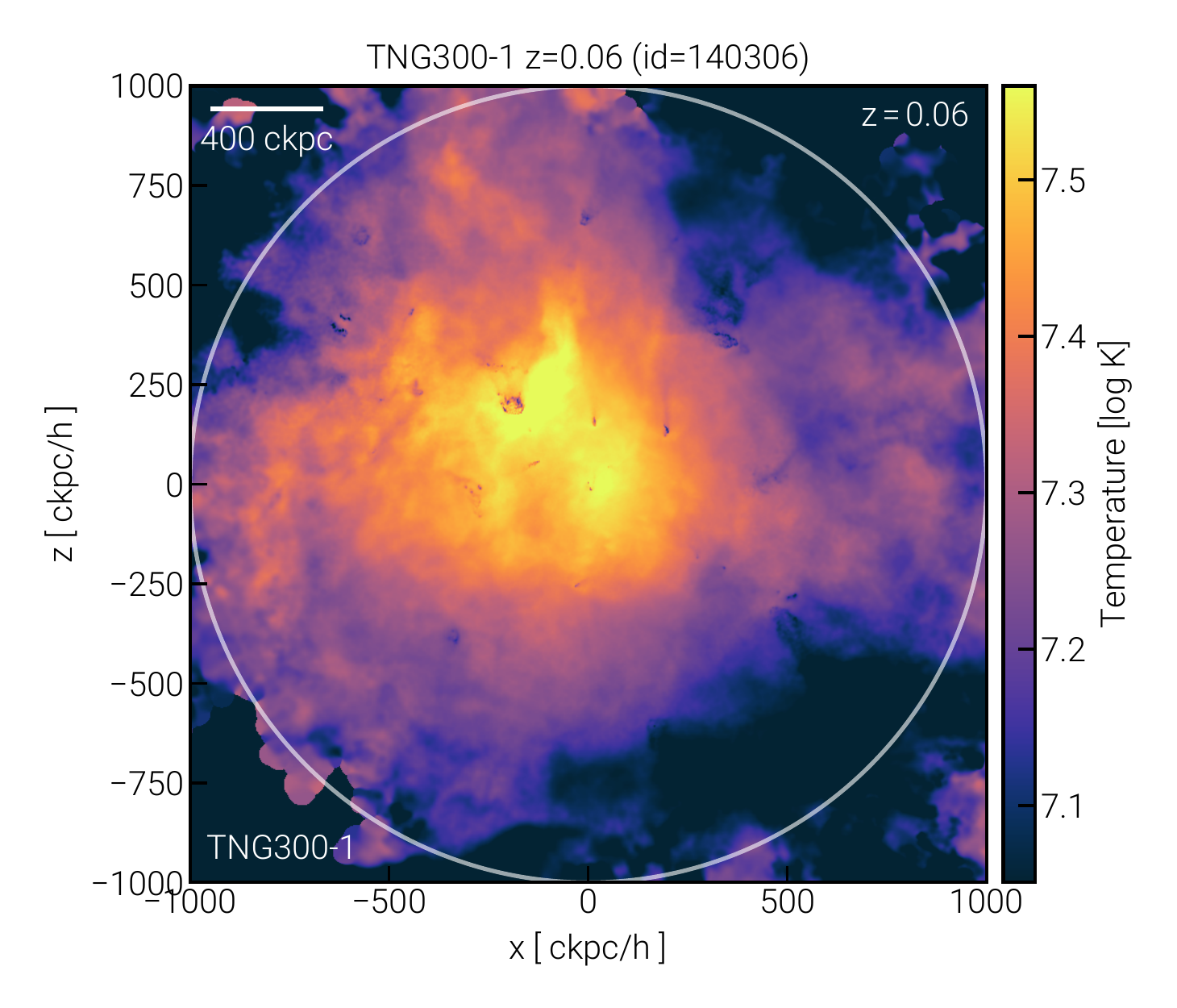}
\\
\includegraphics[width=6.25cm]{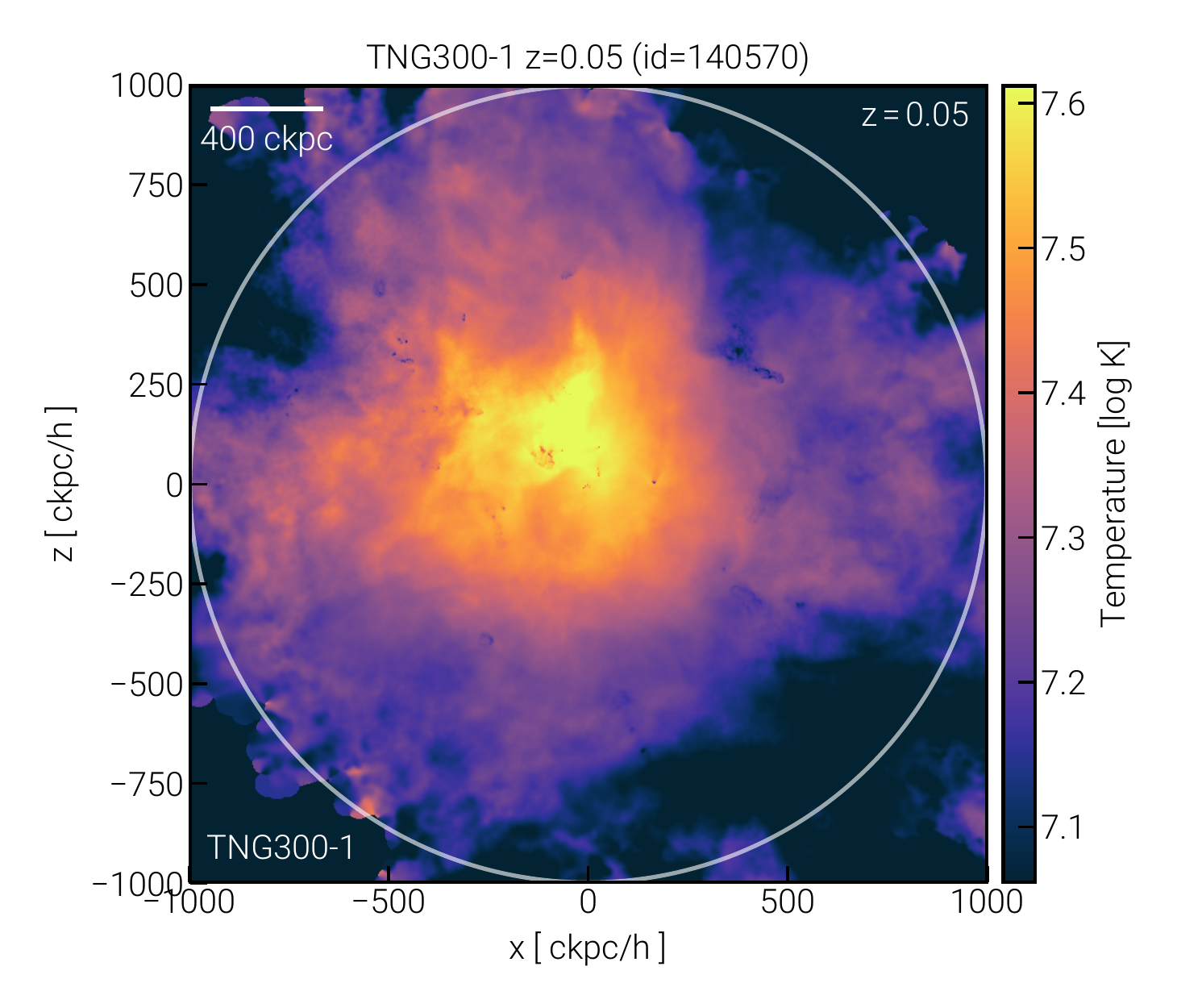}
\hspace{-0.4cm}
\includegraphics[width=6.25cm]{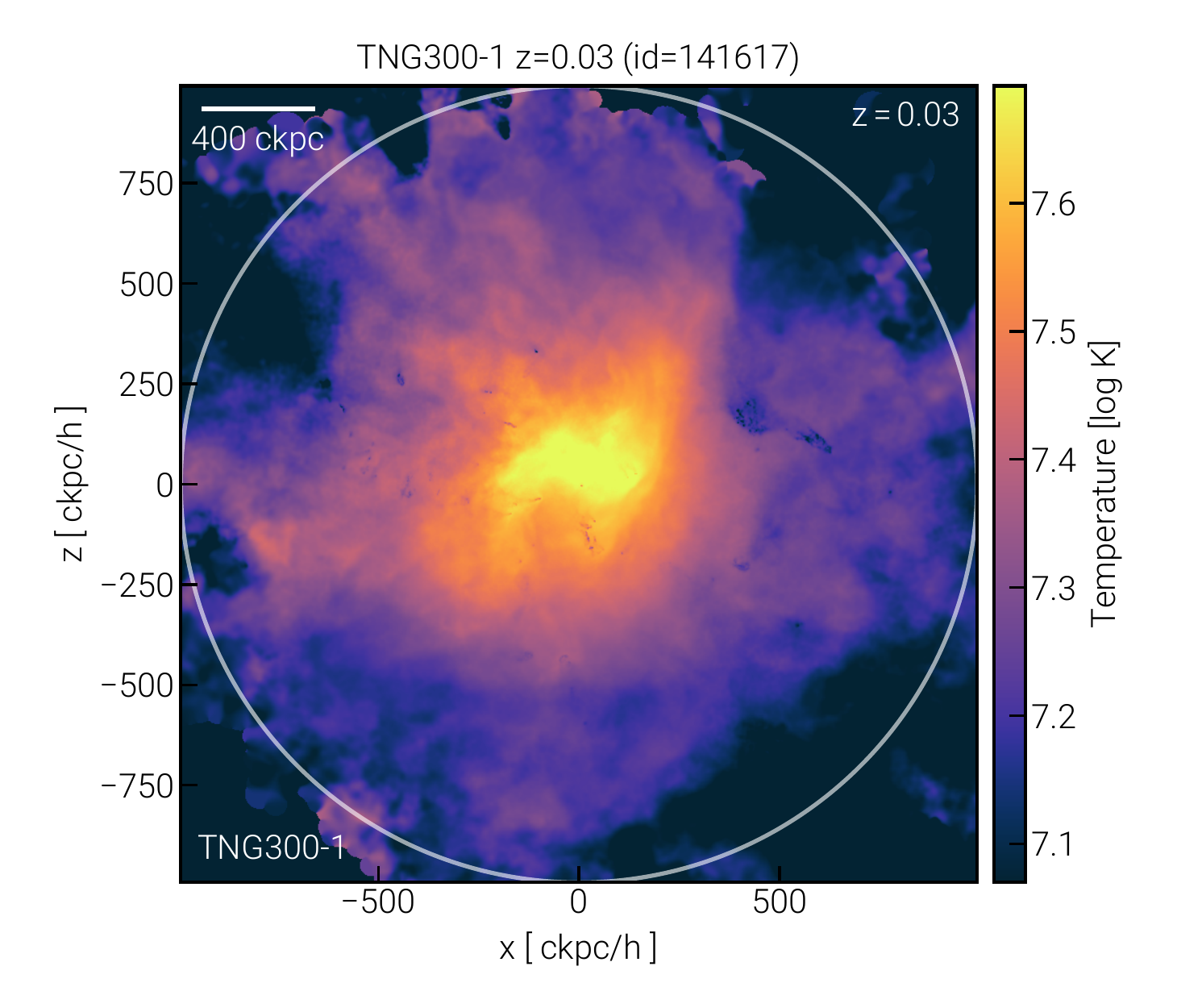}
\hspace{-0.4cm}
\includegraphics[width=6.25cm]{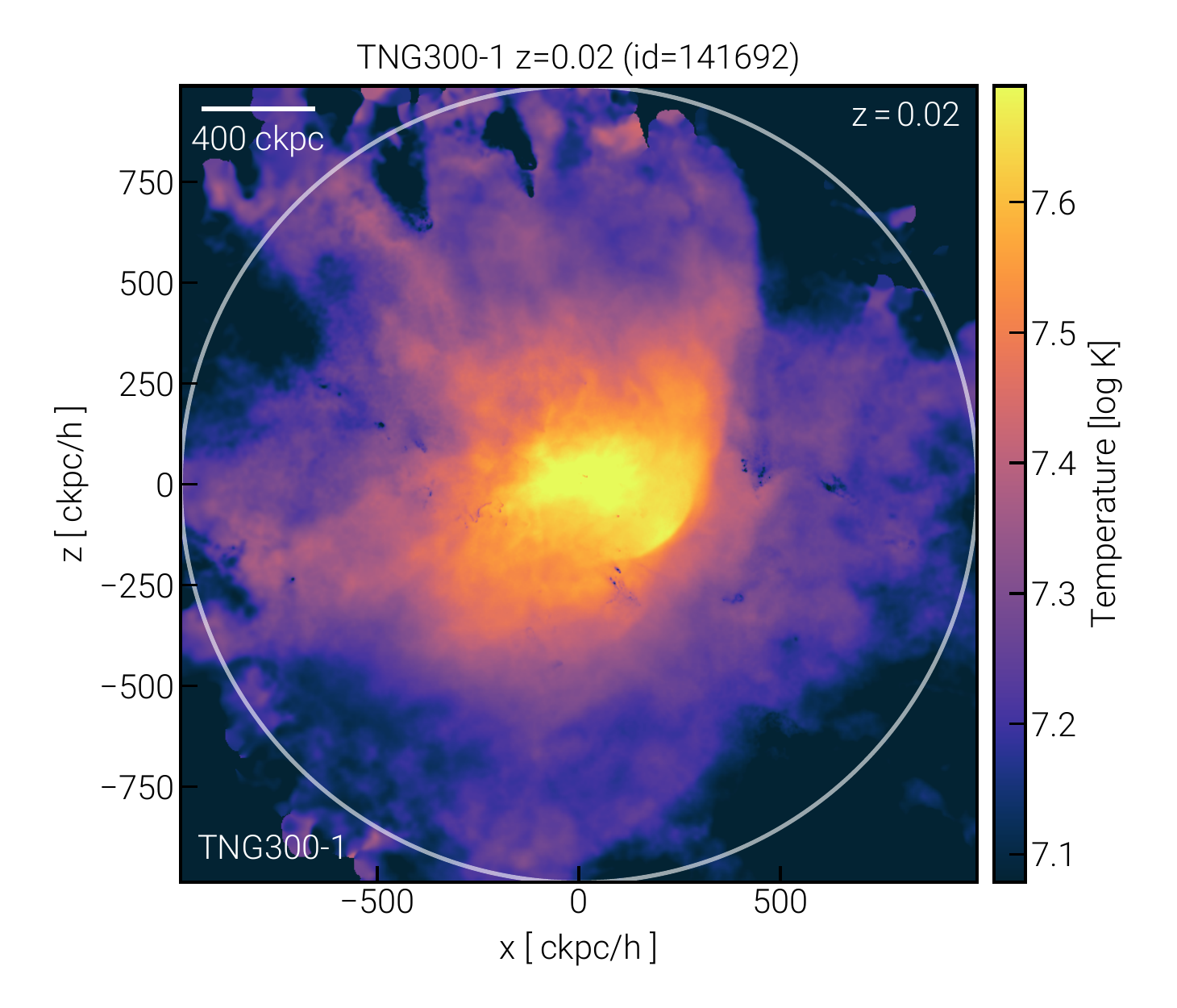}
\caption{Temperature maps of the interacting clusters 148391-148395 at earlier stages of the merger.
The smaller cluster moves from the upper left to the lower right in subsequent images corresponding to
the times between $z=0.08$ and $z=0.02$.}
\label{tempmaps}
\end{figure*}

\begin{figure}
\centering
\includegraphics[width=8.6cm]{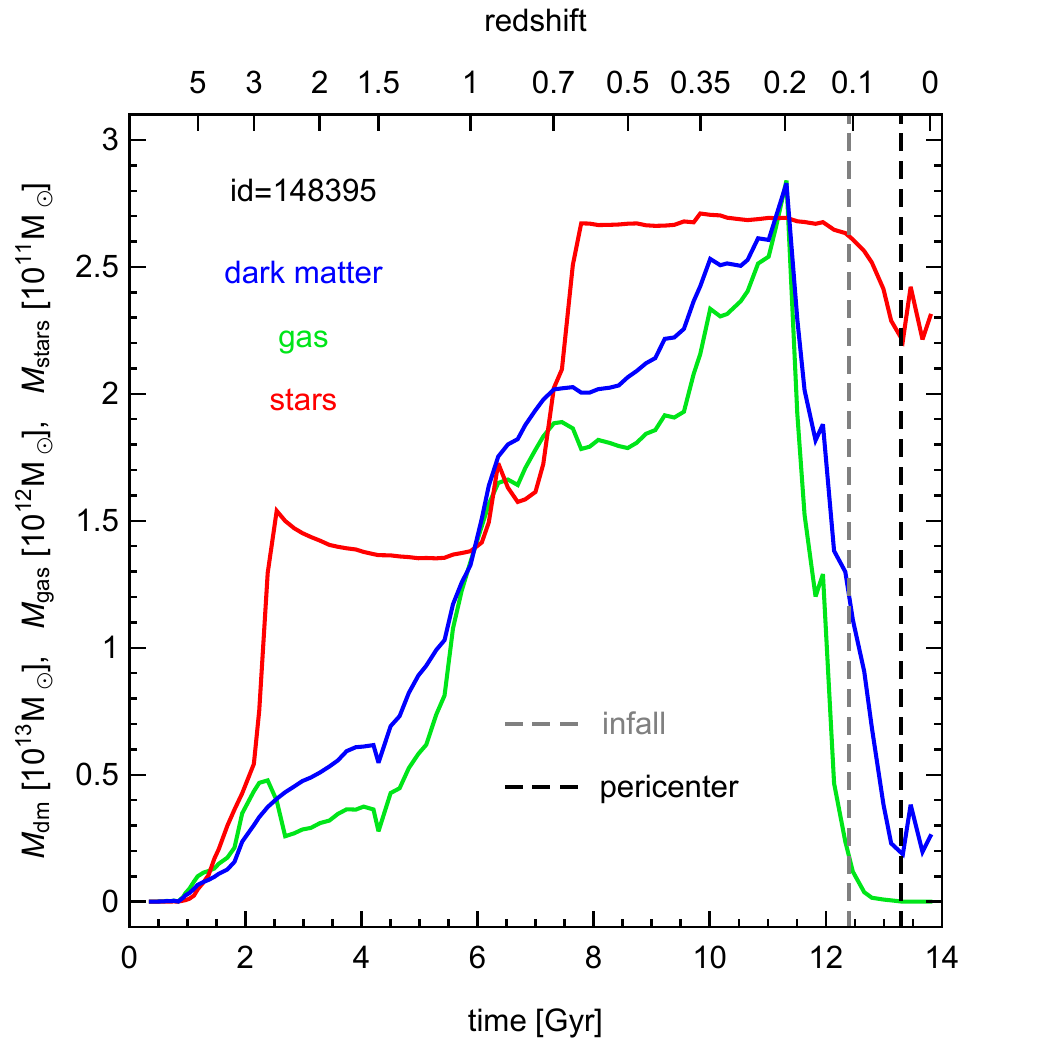}
\caption{Mass evolution of the smaller of the interacting clusters (id = 148395). The blue, green, and red lines
show the measurements for dark matter, gas, and stars, respectively. The units of the different mass components
differ by an order of magnitude. The vertical dashed lines mark the infall time and the time of the pericenter
passage.}
\label{mass}
\end{figure}

In order to trace the interaction of the two clusters in more detail, I followed the evolution of the two subhalos over
the last few simulation snapshots. Since the bow shock is best visible in the temperature maps, in Fig.~\ref{tempmaps}
I show different stages of the interaction of the same system in terms of this property in earlier outputs of the
simulation, from $z=0.08$ ($t = 12.7$ Gyr) to $z=0.02$ (13.5 Gyr). Starting in the upper-left panel of the figure, the
smaller cluster is still clearly visible as a distinct structure, and in the subsequent images, it can be seen to
approach the bigger cluster and pass it, forming a bow shock that starts to be visible in the lower-right panel
corresponding to $z=0.02$.

Tracing the positions and velocities of the two structures in time, one can determine the approximate relative orbit of
the two interacting subhalos. It turns out that the smaller cluster passed close to the bigger one
only once in their history, about 0.5 Gyr ago. This is to be expected, as the smaller structure would otherwise not
bring enough gas to form a clear bow shock. The smaller cluster passed close to the center of the bigger one with a
pericenter of about 190 kpc and with velocity of 1670 km s$^{-1}$.

The mass of the bigger cluster (id = 148391) at $z=0$ is $(2.7, 0.49, 0.015) \times 10^{14}$ M$_{\odot}$ in the dark
matter, gas, and stellar component, respectively, and these values grew a little over the last few simulation
snapshots. In contrast, the smaller cluster loses a dominant fraction of its mass during the interaction. The
mass evolution of the different components of this object is shown in Fig.~\ref{mass}. One can see that at $t = 11.3$
Gyr ($z=0.2$), the subcluster had a maximum mass in all components, reaching $(2.8, 0.28, 0.027) \times 10^{13}$
M$_{\odot}$ in the dark matter, gas, and stars, respectively, so it was about ten times less massive than the main
cluster. After the interaction, however, the mass assigned to it dropped greatly. It lost all its gas and about 90\% of
its dark mass, while the stars were much less affected. These values signify a particularly strong case of tidal and
ram-pressure stripping resulting from the interaction.

\begin{figure}
\centering
\includegraphics[width=9cm]{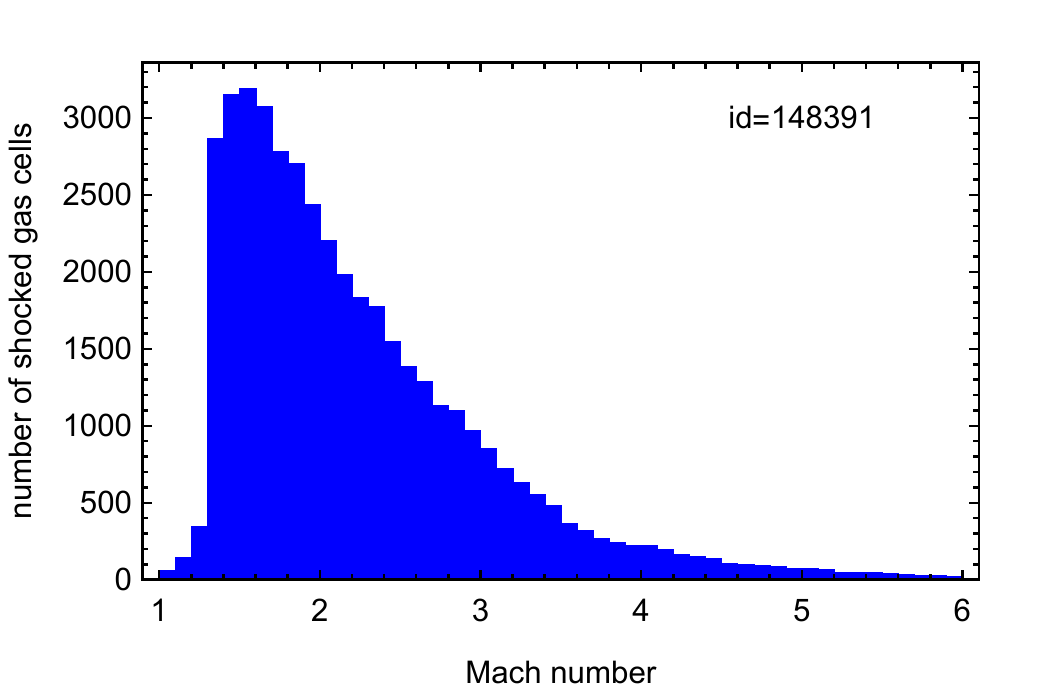}
\caption{Distribution of Mach numbers of shocked gas cells in cluster with id = 148391.}
\label{histogram148391}
\end{figure}

\begin{figure*}
\centering
\includegraphics[width=6.25cm]{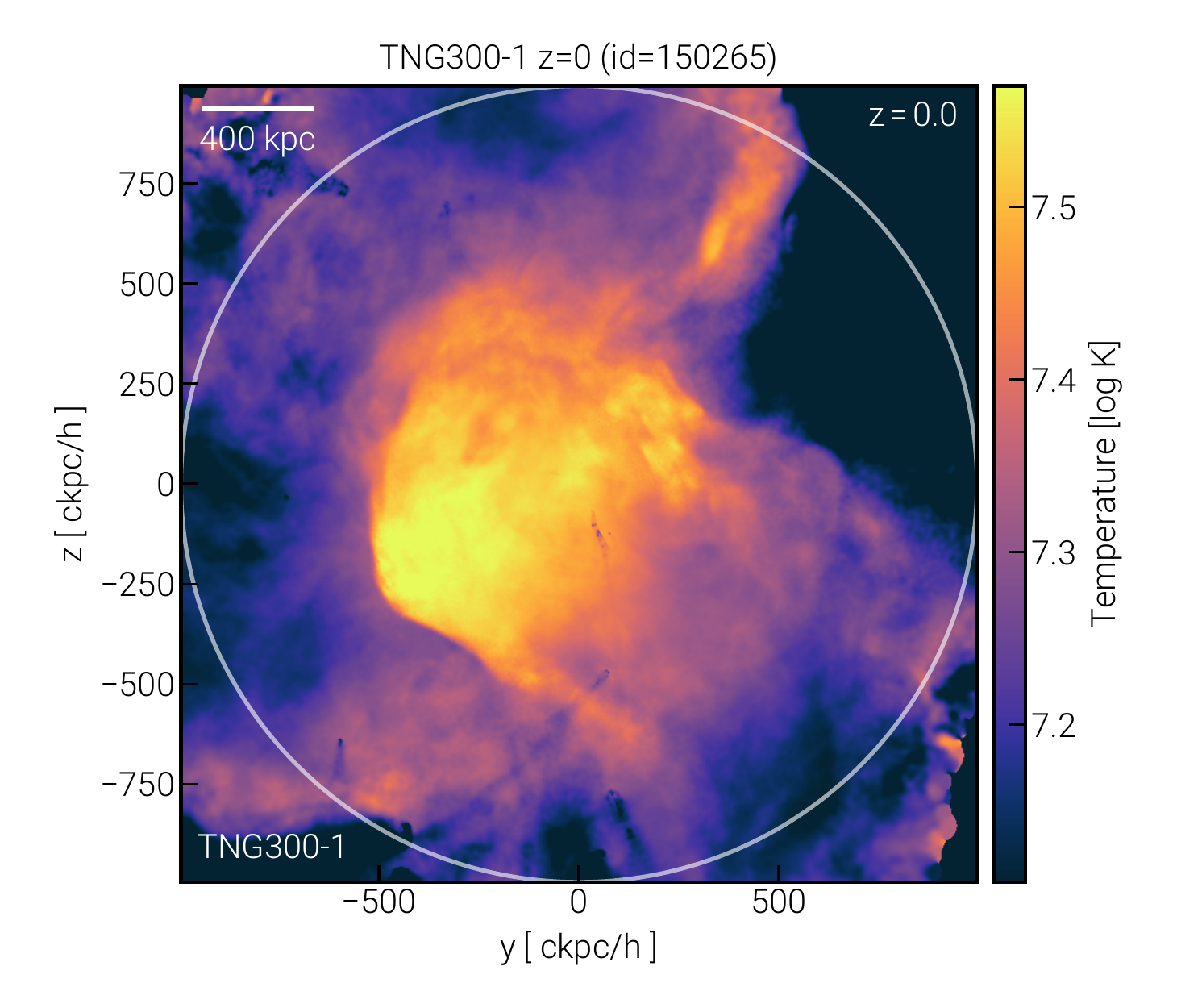}
\hspace{-0.4cm}
\includegraphics[width=6.25cm]{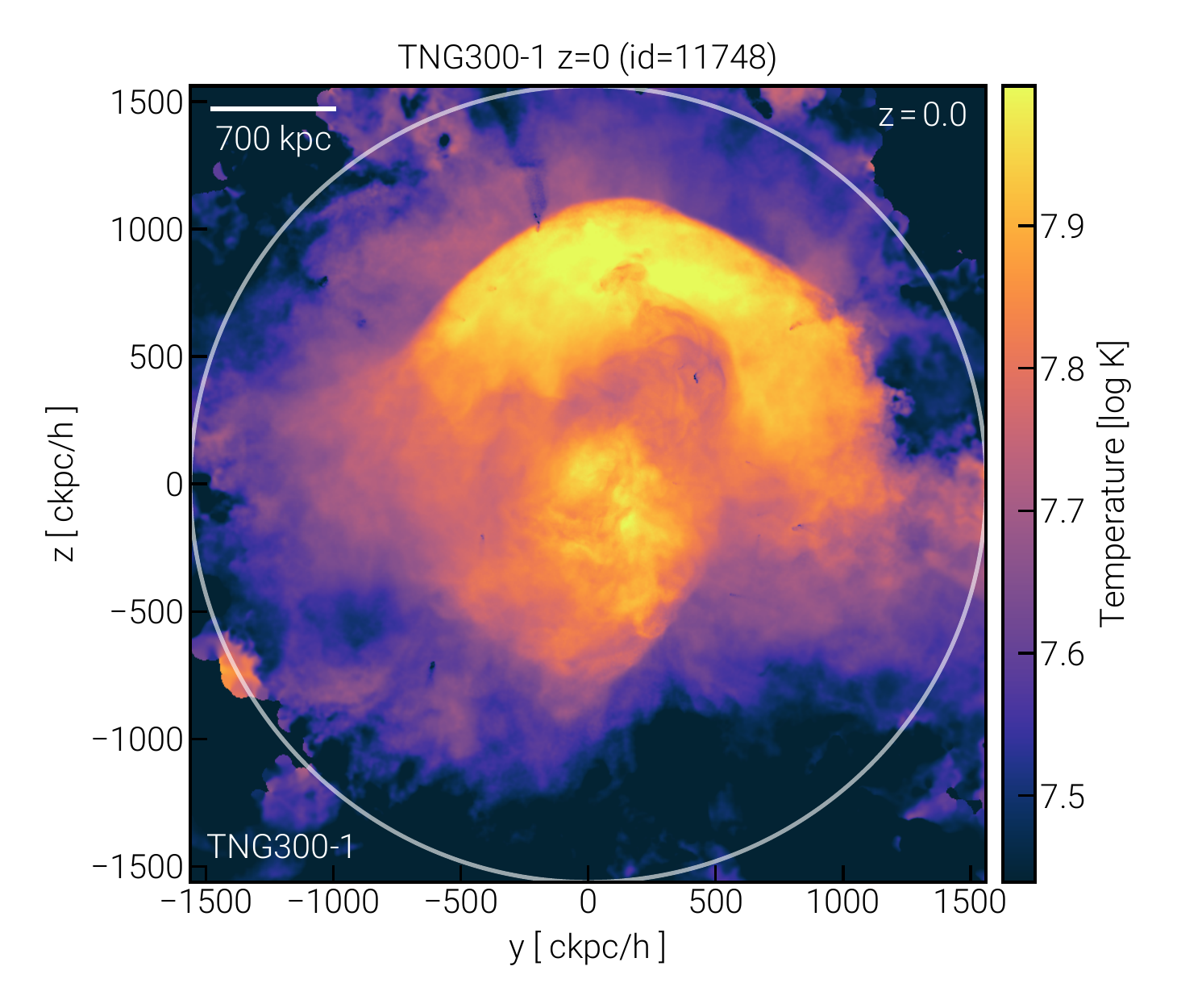}
\hspace{-0.4cm}
\includegraphics[width=6.25cm]{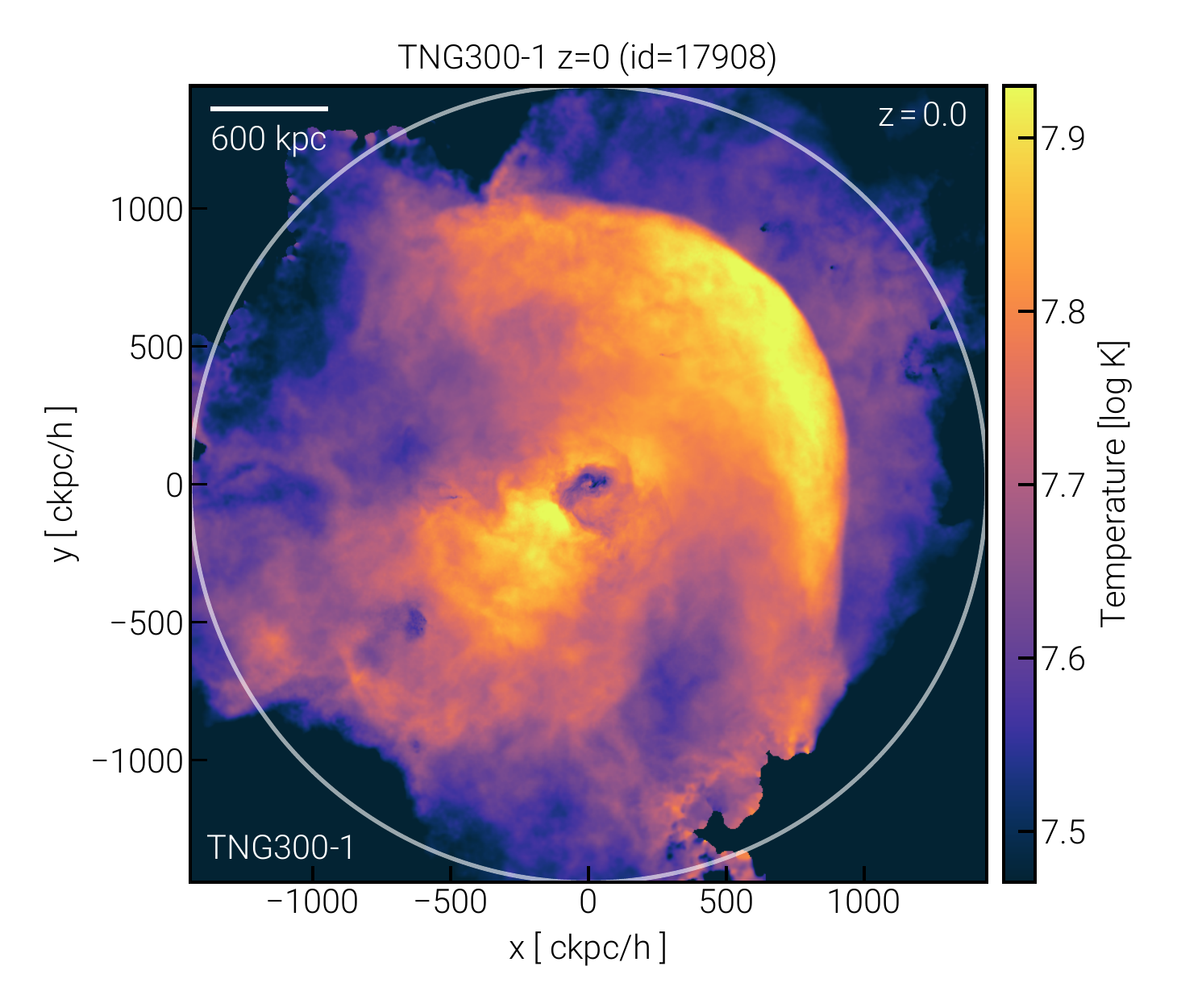}
\\
\includegraphics[width=6.25cm]{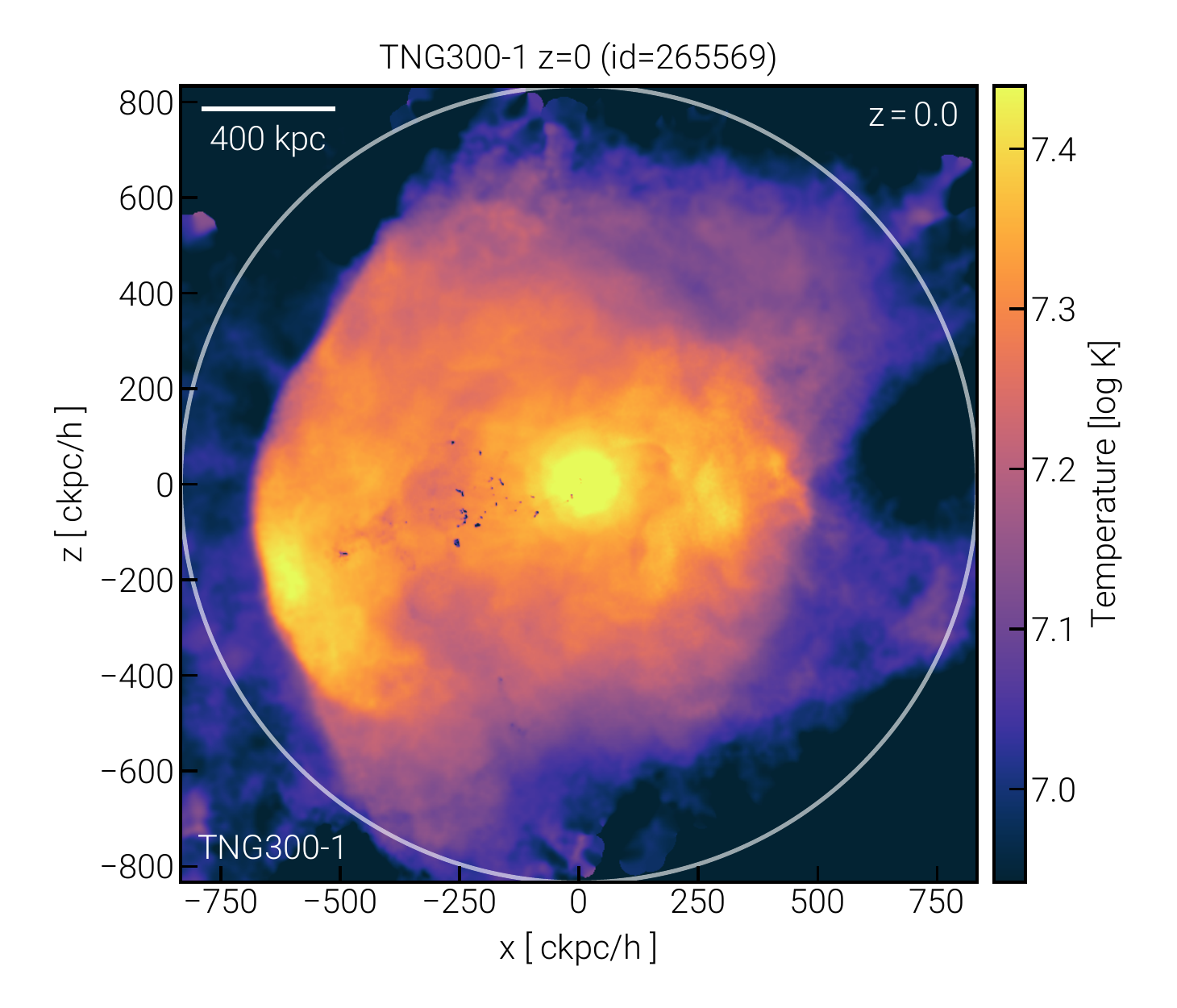}
\hspace{-0.4cm}
\includegraphics[width=6.25cm]{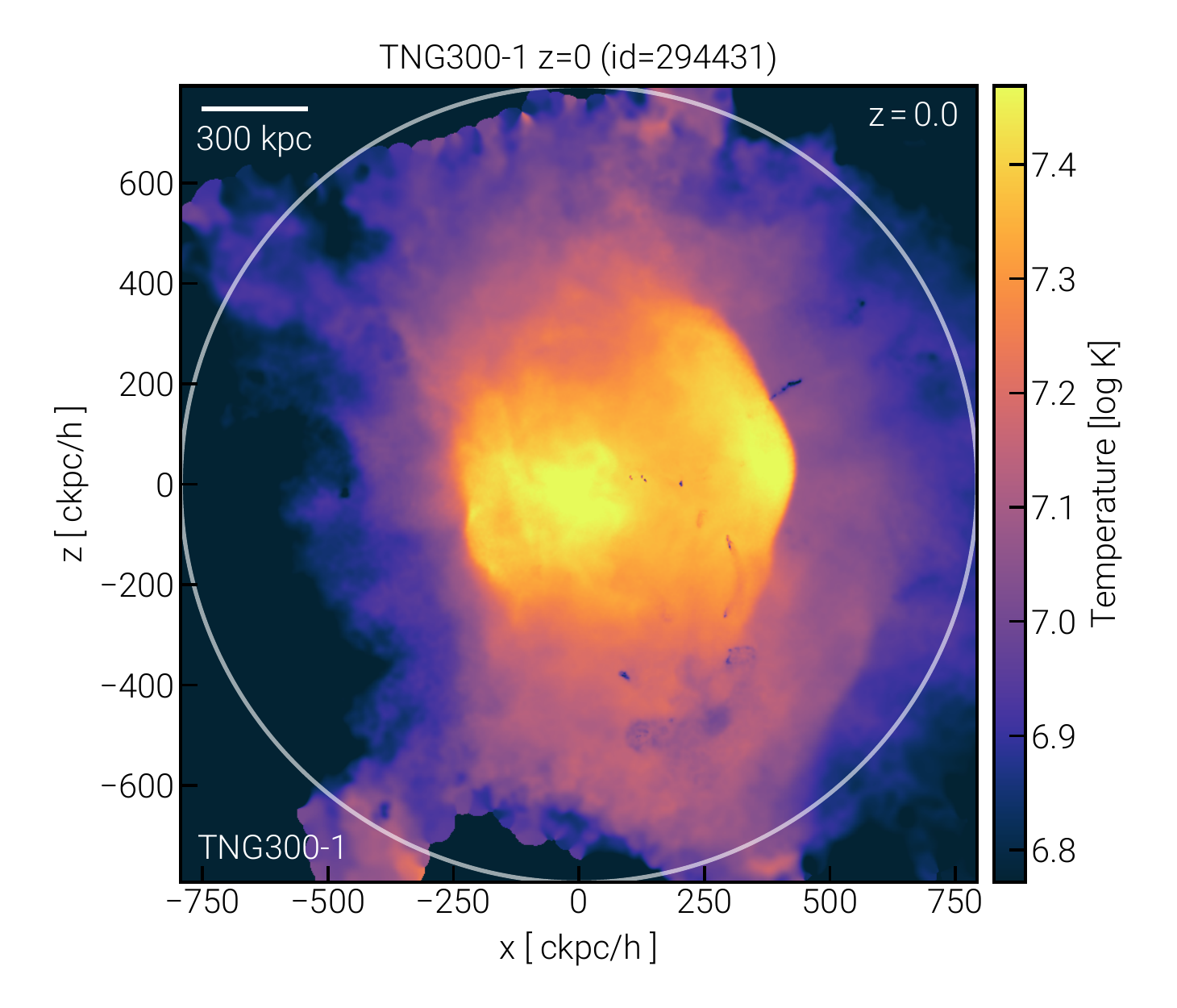}
\hspace{-0.4cm}
\includegraphics[width=6.25cm]{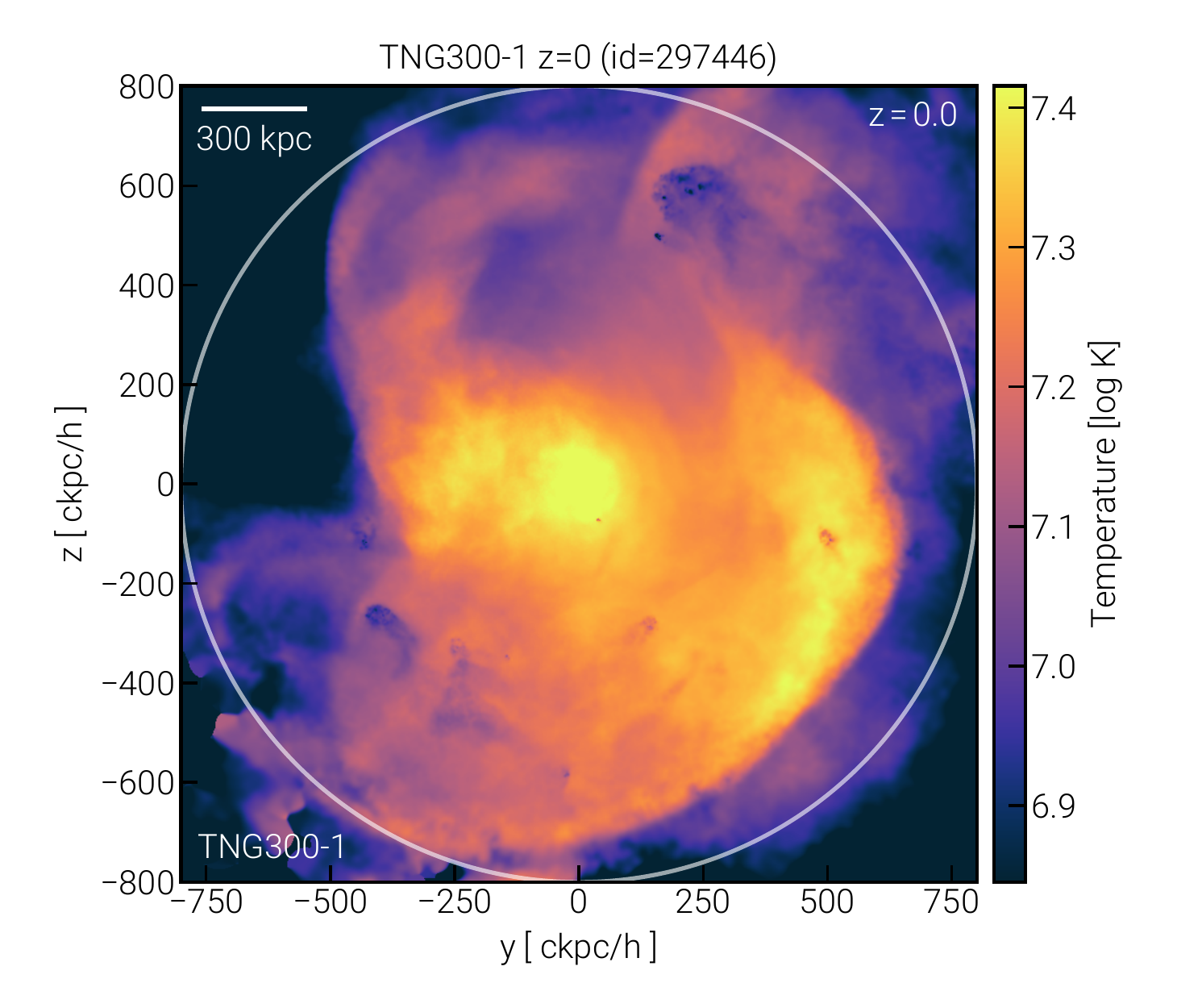}
\caption{Temperature distribution maps of six more examples of interacting clusters with regular single bow
shocks at $z=0$.}
\label{sixmore}
\end{figure*}

\begin{table*}
\caption{Properties of selected merging clusters from IllustrisTNG300.}
\label{properties}
\centering
\begin{tabular}{c c c c c c c c c c c }
\hline\hline
Main cluster & Subcluster &  $M_{\rm main,tot} (t_0)$ & $t_{\rm max}$ & $M_{\rm sub,tot} (t_{\rm max})$ & $t_{\rm peri}$  & $d_{\rm peri}$ & $v_{\rm peri}$ & $f_{\rm sh}$  & Median $\mathcal{M}$\\
id           & id         & [$10^{13}$ M$_\odot$]     & [Gyr]         & [$10^{13}$ M$_\odot$]           & [Gyr]           & [kpc]          & [km s$^{-1}$   &               &               \\ \hline
148391       & 148395     &  32.1                     &  11.3         &  3.14                           &  13.3           &  190           &  1670          & 0.010         &  2.03         \\
150265       & 150266     &  33.1                     &  10.0         &  3.80                           &  13.3           &  170           &  1820          & 0.008         &  1.97         \\
 11748       &  11750     & 131.0                     &  10.8         &  9.17                           &  13.2           &  530           &  3070          & 0.005         &  1.92         \\
 17908       &  17909     & 110.4                     &  10.0         &  6.22                           &  13.0           &  140           &  3430          & 0.005         &  1.89         \\
265569       & 265570     &  20.1                     &  10.8         &  1.62                           &  13.0           &  220           &  1530          & 0.007         &  2.00         \\
294431       & 294432     &  17.5                     &   9.8         &  3.24                           &  13.3           &  250           &  1110          & 0.011         &  2.02         \\
297446       & 297447     &  17.2                     &   7.8         &  1.80                           &  12.9           &  200           &  1390          & 0.009         &  1.97         \\
 58081       &  58083     &  55.0                     &  13.7         & 56.2                            &  13.3           &  260           &  1363          & 0.007         &  1.97         \\
248325       & 248326     &  20.6                     &  12.8         & 17.3                            &  13.5           &  140           &   860          & 0.012         &  1.83         \\
359091       & 359092     &  14.1                     &  13.3         & 14.1                            &  13.1           &  100           &  1270          & 0.010         &  2.17         \\
\hline
\end{tabular}
\end{table*}

Finally, Fig.~\ref{histogram148391} describes the properties of the shocked gas in the merging clusters in
terms of the distribution of Mach numbers of shocked gas cells (with Mach number $\mathcal{M} > 1$) present in the
cluster with id = 148391 at $t_0$. These shocked gas cells comprise about 1\% of all the gas in the cluster. One can see
that the distribution is strongly asymmetric, with the maximum below two and a median Mach number of 2.03.

\section{Other examples of merging clusters}

Temperature maps for six more examples of the merging clusters, identified using the procedure described in the
previous sections, are shown in Fig.~\ref{sixmore}. The identification numbers of the main (more massive)
clusters, starting with the flagship case of the previous section, are listed in the first column of
Table~\ref{properties}. The selected merging clusters show a range of values from 0.005 to 0.012 for the fraction of
shocked gas cells in the main cluster, $f_{\rm sh}$. These values are listed in the penultimate column of
Table~\ref{properties}. The last column of the table contains the median Mach numbers, $\mathcal{M}$, of the shocked
gas cells in the main cluster, which all turned out to be close to two, as in the flagship example. The
distributions of the Mach numbers of the other clusters are also similar to the one shown in Fig.~\ref{histogram148391}
for id = 148391.

The second column of Table~\ref{properties} lists the identification numbers of the subclusters taking part in
the mergers, identified using the procedure described in Section~2. Checking the mass evolution of the subclusters
confirmed that in all cases it behaved similarly as in the case of the flagship example. Namely, the subcluster always
lost most of its gas and dark matter, while the stellar core remained essentially intact. The secondary cluster is
identified with a subhalo, which in all cases is a massive red elliptical and a central galaxy of a group. It contains
its own hot gas component but has little cold gas and does not form stars. This gas is stripped very
quickly and assigned to the main cluster.

The third column of the table gives the masses of the main clusters at the present time, $t_0$, which corresponds to
the last simulation output. The fourth column lists the times, $t_{\rm max}$, when the subclusters reached their
maximum masses, which are given in the fifth column. Comparing the masses of the interacting clusters listed in rows
one to six of Table~\ref{properties}, one can see that the pair in the second row of the table, with the main cluster
id = 150265, is almost identical to the flagship example in the first row. The two objects in rows three and
four of the table, id = 11748 and id = 17908, have the most massive main clusters in the sample, with present total
masses $M_{\rm main,tot} (t_0) > 10^{15}$ M$_\odot$. The next two main clusters in rows five and six have an order of
magnitude smaller masses of $\sim 2 \times 10^{14}$ M$_\odot$.

\begin{figure}
\centering
\includegraphics[width=8.4cm]{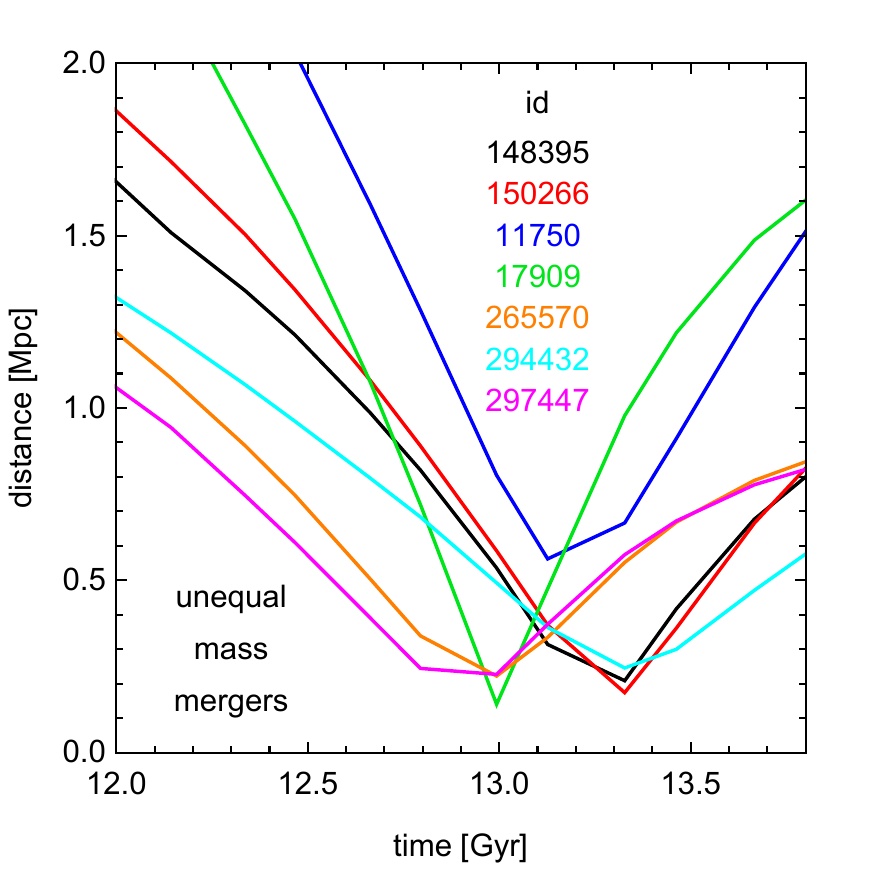}
\includegraphics[width=8.4cm]{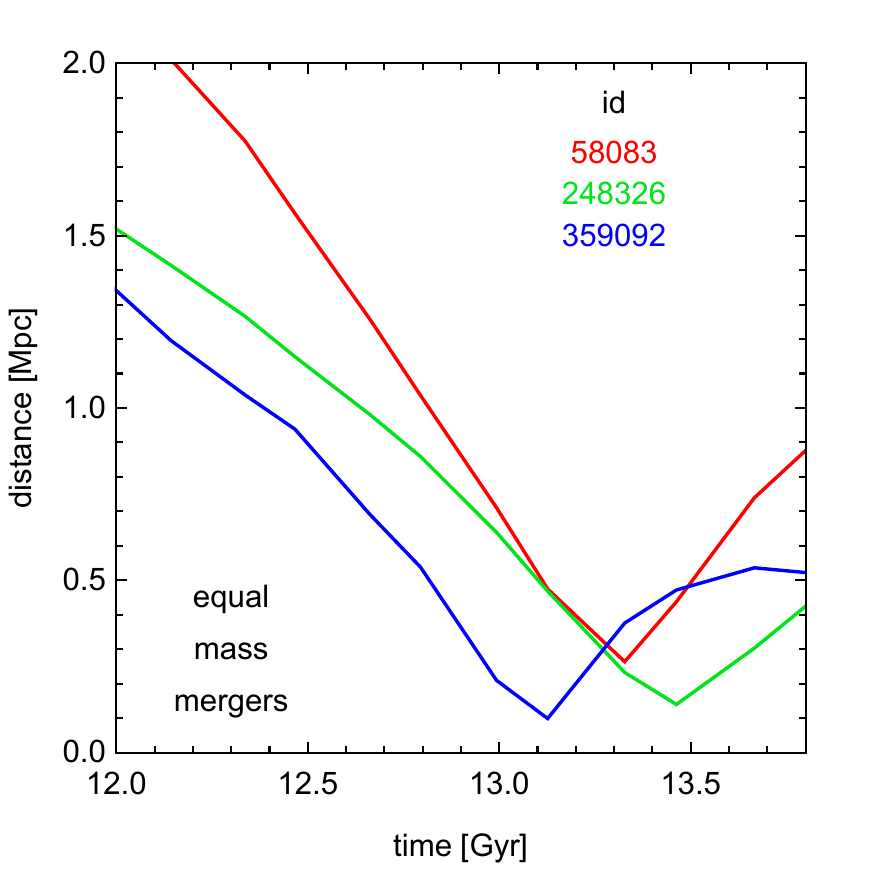}
\caption{Distance of the subcluster to the main cluster as a function of time during the last 1.8 Gyr of evolution.
Colors correspond to different subclusters with id numbers (at $z=0$) given in the legend. The upper panel shows the measurements for unequal-mass mergers with single bow shocks. The lower panel shows the measurements for equal-mass mergers with two
similar bow shocks.}
\label{distances}
\end{figure}

\begin{figure*}
\centering
\includegraphics[width=6.25cm]{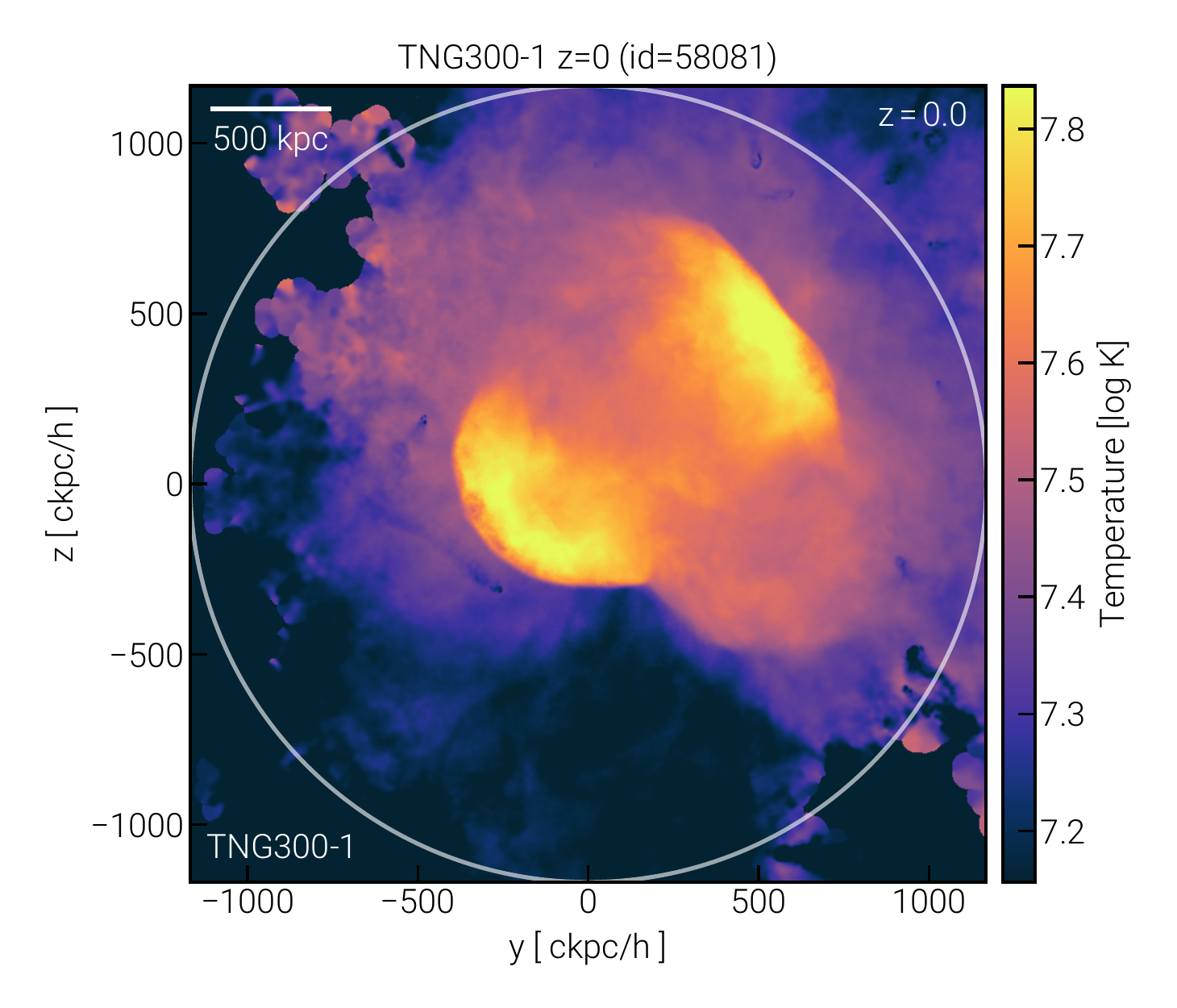}
\hspace{-0.4cm}
\includegraphics[width=6.25cm]{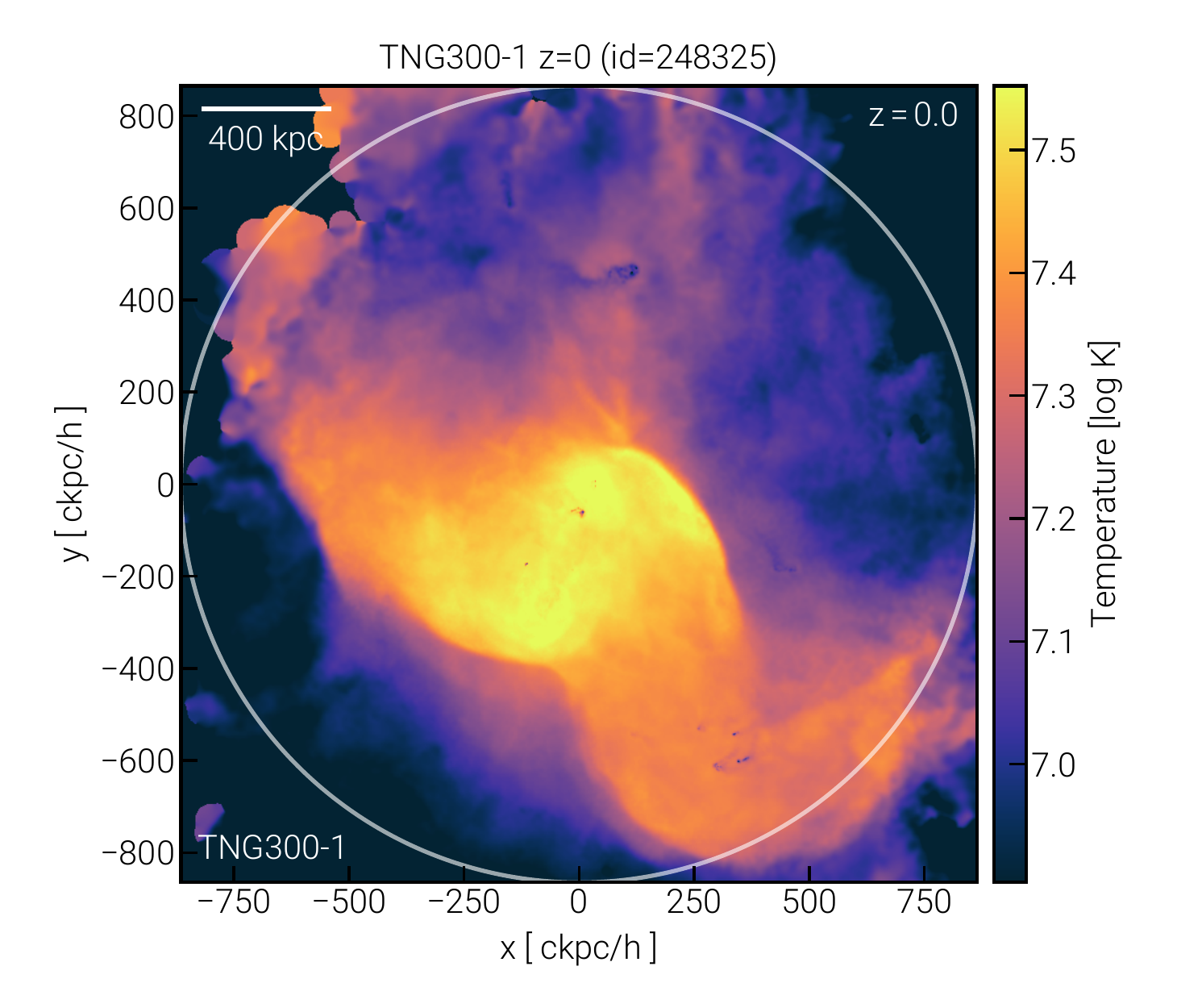}
\hspace{-0.4cm}
\includegraphics[width=6.25cm]{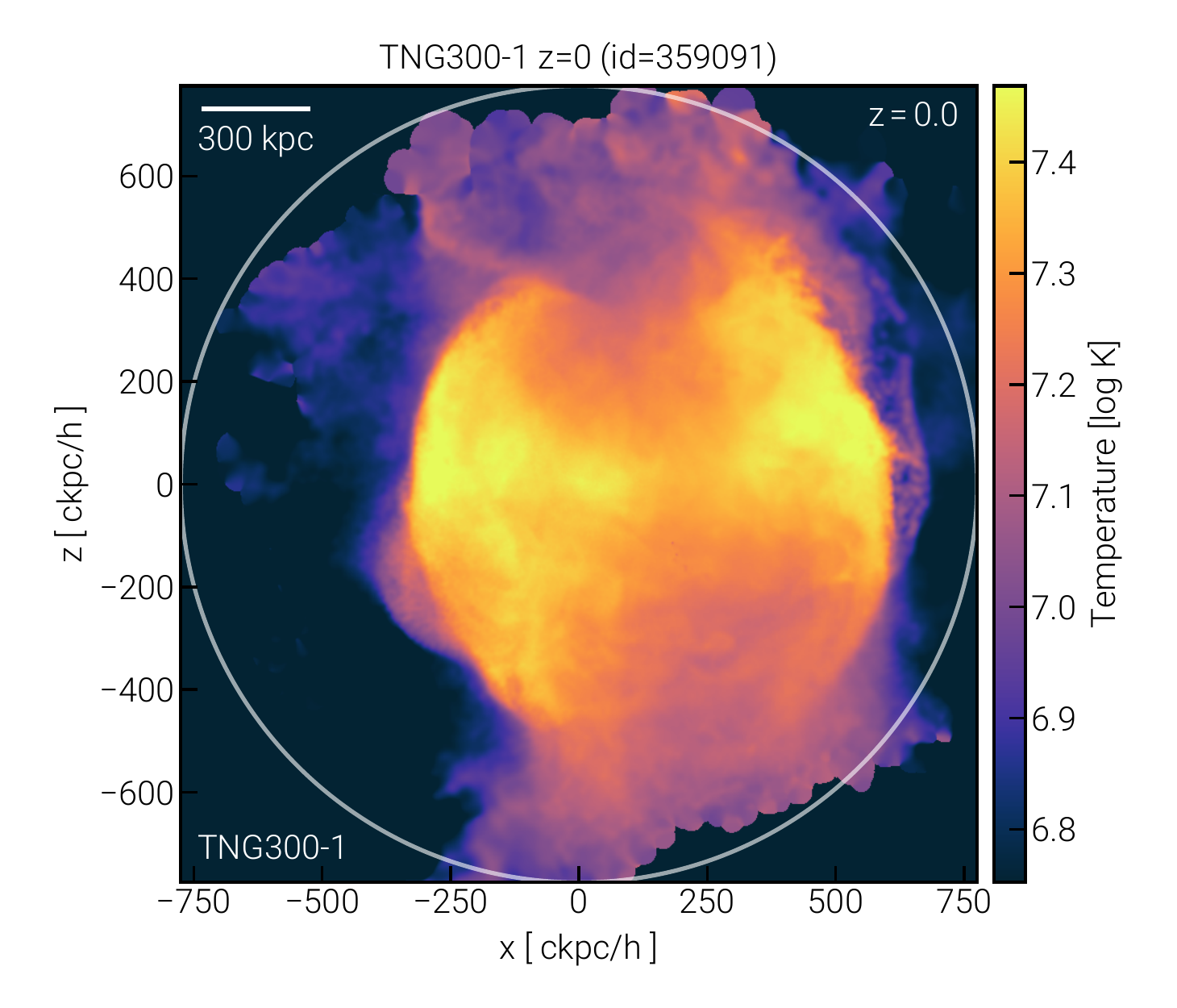}
\caption{Temperature distribution maps of three examples of interacting clusters with comparable mass at $z=0$.}
\label{equalmass}
\end{figure*}

Interestingly, by studying the orbits of the subclusters with respect to the main clusters, I found that they all behaved
in a manner similar to the flagship example. Namely, the subclusters interacted with the bigger cluster only once (i.e., they had a
single pericenter passage around it) between 0.9 and 0.3 Gyr ago. The sixth column of Table~\ref{properties} lists the
times of the pericenter of the interaction ($t_{\rm peri}$), that is, the times when the two clusters were at their
closest approach. The next two columns list the distances, $d_{\rm peri}$, and relative velocities, $v_{\rm
peri}$, of the two clusters at the pericenter. The values of these orbital parameters were obtained by interpolation
between the available simulation outputs that are separated by rather large time steps of 0.1-0.2 Gyr at these times.
The numbers were given with the expected accuracy of 0.1 Gyr, 10 kpc, and 10 km s$^{-1}$ for the pericenter times,
distances, and velocities, respectively.
In the upper panel of Fig.~\ref{distances}, the orbits of the subclusters in terms of their distance from the
main cluster as a function of time in the last 1.8 Gyr are plotted. The pericenter distances most often are on the
order of 200 kpc, but they vary between 140 kpc (id = 17909) and 530 kpc (id = 11750). These two objects also
approach their pericenters with the greatest velocities, exceeding 3000 km s$^{-1}$, probably because their main
clusters are the most massive. In particular, the relative velocity of the 17908-17909 pair at the pericenter is 3430
km s$^{-1}$. The remaining subclusters have velocities below 2000 km s$^{-1}$, similar to the flagship example of id =
148395.

\section{Equal-mass mergers}

Another subclass of merging clusters I discovered among the 200 most massive clusters of IllustrisTNG300 is one
where both merging objects are of similar mass. The temperature maps of three such cases, in projection where the bow
shocks are best visible, are shown in Fig.~\ref{equalmass}. All three panels display pairs of similar bow shocks
created by both clusters. The parameters of the clusters taking part in these interactions are listed in the last three
rows of Table~\ref{properties}. These properties are in general similar to the ones for the clusters described in the
previous sections, except for the mass evolution.

Because the masses of the two interacting structures are quite similar, the Subfind algorithm \citep{Springel2001}
assigned nearly all of the total mass of the two structures to only one of them in each simulation snapshot, which
resulted in strong mass variation for a given subhalo in time. The object I identified as the main cluster in
Table~\ref{properties} is the one that was more massive at the end of the simulation, but this was not necessarily the
case during the whole evolution. In fact, the opposite can be true, which is the case for the 58081-58083 pair (eighth
row of Table~\ref{properties}). The final mass of id = 58081 is $5.5 \times 10^{14}$ M$_\odot$, but the maximum mass of
the progenitor of id = 58083 at $t_{\rm max}$ is $5.6 \times 10^{14}$ M$_\odot$. Thus, estimating the mass ratio of the
two structures at any given snapshot would lead to erroneous results. Instead, they should be considered as equal-mass
mergers, and the mass per cluster should be estimated as half of any of the two objects.

The relative motion of the two interacting clusters in each equal-mass pair is illustrated in the lower panel of
Fig.~\ref{distances} in terms of the distance between them as a function of time. The pericenter distances and times
turned out to be similar to those of the unequal-mass mergers of the upper panel, although they include two extreme
cases. First is the interaction of the 248325-248326 pair, which took place the most recently of all the examples, only
0.3 Gyr ago. The event manifests itself
in a small distance between the two bow shocks in the middle panel of Fig.~\ref{equalmass}. Second case is the pair
359091-359092. These clusters interacted with the smallest pericenter distance, only about 100 kpc. One may also note
that the 58081-58083 pair approached each other more quickly, probably because of their larger masses.

\section{Discussion}

I studied examples of merging clusters selected from the last output of the IllustrisTNG300 simulation. All the merging
events detected in this study bear strong resemblance to each other in many respects. I found that in order to produce
a clearly visible bow shock in the temperature map within the virial radius of the main cluster, the cluster
interaction must have taken place relatively recently, between 0.9 and 0.3 Gyr ago. The subclusters passed close to the
bigger objects only once, with a typical pericenter on the order of 200 kpc and a velocity on the order of 2000 km
s$^{-1}$. Only in such a configuration does the subcluster contribute enough gas to shock a substantial fraction of the
gas content in the system. These shocked gas fractions are on the order of 1\% in all cases, and the median Mach
numbers of shocked gas cells are always close to two. This means that the shocks I identified in these merging clusters
are internal, according to the classification of \citet{Ryu2003}. In other words,  the shocks originate from an already
preheated gas and result in Mach numbers typically below ten \citep{Schaal2015}. These results should not depend
strongly on resolution if the tests preformed by \citet{Schaal2016} on the Illustris simulations also hold for the
IllustrisTNG, as the authors found that the positions of the shocks and the measured Mach numbers were similar between
simulations of different resolutions.

The flagship example among the interacting clusters seems to be representative of the whole sample discussed in this
paper, and I studied it in more detail. Other identified cases can be tentatively divided into two
subsamples of unequal- and equal-mass mergers, with the former characterized by a single well-defined bow shock and the
latter by two similar shocks in the temperature maps. The presented examples are not the only ones with bow shocks in
the adopted mass range; they are only the most clear, binary ones. There are others, such as id = 225690, which has two
asymmetric bow shocks and is thus an intermediate case between the unequal- and equal-mass mergers, which have
either a single shock or two symmetric bow shocks, respectively. Taking the number of
identified clear cases at face value, one could estimate the fraction of merging clusters in the mass range above $1.4
\times 10^{14}$ M$_{\odot}$ to be at least 10/200, that is, 5\%. Taking into account the volume of the simulation box,
this would correspond to the lower limit on the number density of merging clusters in this mass range of $3.6 \times
10^{-7}$ Mpc$^{-3}$. It is difficult to compare these numbers with real data since many cluster mergers may remain
undetected in X-ray observations and the exact dynamical status of observed interacting groups is
most often impossible to determine \citep{Tempel2017}.

I selected the merging clusters through visual inspection of temperature maps generated using the visualization tools
provided by the IllustrisTNG team. One might expect that a more straightforward and reliable way to identify such
objects would be to select clusters with high fractions of shocked gas. However, I found that there is no one-to-one
correspondence between the clusters with the highest fractions of shocked gas and those with the best visible bow
shocks. In other words, not all the clusters with the highest shocked gas fractions display bow shocks, and these fractions among
the clusters with bow shocks are not the highest, although they are above the median.

This lack of correspondence between the merging clusters and those with the biggest shocked gas fractions may be due to
at least two reasons. First, the number of gas cells actually assigned to a cluster strongly depends  on
the environment. If there are massive structures nearby, the gas mass content of the cluster will naturally
be truncated by the Subfind algorithm, and its total gas mass will be smaller. Second, the shocks occurring in the gas can be due to not only mergers between large structures but also the result of active galactic nucleus (AGN) feedback
\citep{Ubertosi2023}. I verified, however, that the significant AGN feedback was present in the main clusters
and subclusters only at higher redshifts of $z = 2-3$, so it does not contribute to the creation of large-scale bow
shocks at $z=0$. Therefore, I can confidently claim that all the large-scale bow shocks identified in this study are
associated with merging clusters.

One of the interesting properties of the merging clusters is their mass ratio. The mass ratios of the merging
structures are not straightforward to estimate in observations nor in simulations. Although I selected the cases of
merging clusters from the last simulation output, they cannot be calculated at the present time from the IllustrisTNG
data because the subcluster is already stripped of most of its mass. For the subcluster mass, it thus seems reasonable
to assume its maximum mass value in the entire history.
The times of this maximum mass for the unequal-mass mergers turned out to be on the order of $t_{\rm max} = 10$ Gyr,
which is about three Gyr prior to the typical time of the pericenter passage. The masses of the main clusters have not
dramatically changed since the interaction, so the mass ratios can be estimated from the values given in
Table~\ref{properties}. I therefore find that the mass ratios of the unequal-mass mergers vary between five and 18,
with the most typical values being on the order of ten. A similar approach can be used in the case of equal-mass
mergers. For these three cases, $t_{\rm max}$ is larger due to the
changes in the mass assigned to the two similar structures, but it is not actually significant.

A feature of the real bullet cluster that attracted considerable attention (because of its significance in the
estimates of dark matter distribution) is the offset between the gas associated with the smaller cluster and its dark
matter and galaxies such that the gas seems to be delayed with respect to collisionless components. I have
checked if a similar displacement is present in the simulated clusters; although in the cases studied here, the
subcluster does not retain any well-defined gas component. However, a similar displacement could be observed between
the position of the bow shock and the collisionless component of the subcluster, as demonstrated by Fig. 7 in
\citet{Springel2007}. I measured its value by comparing the distribution of the shocked gas, which forms a thin surface
in 3D, and the position of the center of the remnant stellar core of the subcluster. Among the seven unequal-mass
mergers, four have the subcluster remnant lagging behind the bow shock, typically by a few tens of
kiloparsecs. In two cases, the subcluster core lies within the thickness of the shocked gas layer,
and only in one case (id = 17909) is the subcluster core ahead of the bow shock. It is worth noting that this
is also the subcluster taking part in the collision with the highest speed and that it has a very small pericenter. The
variety of configurations of the displacement found here is in agreement with earlier studies that used controlled
simulations, where it was demonstrated that the displacement is sensitive to
the relative concentrations and gas fractions of the merging clusters and highly time dependent \citep{Springel2007}.

A detailed comparison of the findings presented here to the observed merging clusters is beyond the scope of
this paper. It might be worthwhile, however, to comment on the similarity of the simulated clusters to the real bullet
cluster as a reference case. No close match to this real object was found, but at least one candidate can be proposed
among the presented cases to bear some resemblance to this classic example of merging clusters. As mentioned above,
although the bow shocks of all the simulated clusters are similar in appearance to the one of the bullet cluster, none
of the simulated clusters retains enough gas to form a clear bullet-like component with a lower
temperature. For the bullet to survive, it would probably be necessary for the subcluster to possess a cool core, while
among the selected cases only one of the main clusters appears to have this characteristic (id = 17908; upper-right
panel of Fig.~\ref{sixmore}). In terms of the mass of the main cluster, the relative velocity of the interacting
objects, and the displacement between the bow shock and the collisionless components, the interacting pair 17908-17909
seems to provide the closest match to the real bullet cluster.

However, on average, the velocities and the median Mach numbers of the simulated clusters (approximately two) are more similar to
that of the A520 cluster \citep{Markevitch2005} or Coma \citep{Planck2013}. The lower velocities are probably also the
reason why the times since the pericenter, which I found to be in the range of 0.9-0.3 Gyr, are longer than those
for real clusters, typically estimated to be on the order of 0.3-0.1 Gyr \citep{Markevitch2002, Girardi2008,
Girardi2016}. Another possibility is the detection bias; I looked for clear bow shocks within the virial radius of
the main cluster, while the field of view of X-ray observations is usually smaller, at least for nearby
clusters. The method used in this paper was also different than the one used by observers, I relied on the
(well-resolved) temperature maps of simulated clusters, while for real objects such maps are derived by modeling the
spectra, and mergers are detected using maps of X-ray luminosity. Indeed, the creation of detailed temperature maps
from observations could help in comparisons between real and simulated objects.

It might also be of interest to look for similarities between the
simulated merging clusters described here and the well-studied nearby Virgo cluster. While the range of masses of the
simulated objects is similar to that of Virgo, their dynamical status is rather different. Although the Virgo cluster
is believed to be composed of a few groups \citep{Binggeli1987, Mei2007}, these structures are probably infalling toward
the main cluster for the first time, and no signatures similar to bow shocks have been observed in the X-ray data
\citep{Bohringer1994}. I do not claim to have detected all cases of merging clusters in the
IllustrisTNG data, only the most violent ones that produce pronounced bow shocks. There are probably other
ongoing merger events with smaller structures being accreted with lower velocities or at earlier stages.

I emphasize that the lower relative velocities of the simulated merging clusters studied here are expected
due to the size of the simulation box, which is significantly smaller than what would be required to probe the tail of
the pairwise velocity distribution and find an exact analog of the bullet cluster. Still, the smaller box was
necessary to be able to adequately follow all the physical processes involved in the formation and evolution
of clusters. The presented examples demonstrate that mergers between clusters in configurations less extreme than the
bullet cluster can be satisfactorily reproduced in IllustrisTNG.

\vspace{0.5cm}

\begin{acknowledgements}
I am grateful to the IllustrisTNG team for making their simulations publicly available and to the anonymous
referee for useful comments. Computations for this article have been performed in part using the computer cluster at
CAMK PAN.
\end{acknowledgements}


\begin{thebibliography}{}

\bibitem[{Barnes et al.}(2018)]{Barnes2018} Barnes, D. J., Vogelsberger, M., Kannan, R., et al. 2018, MNRAS, 481, 1809
\bibitem[{Binggeli et al.}(1987)]{Binggeli1987} Binggeli, B., Tammann, G. A., \& Sandage, A. 1987, AJ, 94, 251

\bibitem[{Bohringer et al.}(1994)]{Bohringer1994} B\"{o}hringer, H., Briel, U. G., Schwarz, R. A., et al. 1994,
        Nature, 368, 828

\bibitem[{Bouillot et al.}(2015)]{Bouillot2015} Bouillot, V. R., Alimi, J.-M., Corasaniti, P.-S., Rasera, Y.
        2015, MNRAS, 450, 145
\bibitem[{Chon \& B\"{o}hringer}(2015)]{Chon2015} Chon, G., \& B\"{o}hringer, H. 2015, A\&A, 574, A132
\bibitem[{Clowe et al.}(2004)]{Clowe2004} Clowe, D., Gonzalez, A., Markevitch, M. 2004, ApJ, 604, 596
\bibitem[{Clowe et al.}(2006)]{Clowe2006} Clowe, D., Brada\v{c}, M., Gonzalez, A. H., et al. 2006, ApJ, 648, L109
\bibitem[{Dacunha et al.}(2022)]{Dacunha2022} Dacunha, T., Belyakov, M., Adhikari, S., et al. 2022, MNRAS, 512, 4378
\bibitem[{Girardi et al.}(2008)]{Girardi2008} Girardi, M., Barrena, R., Boschin, W., \& Ellingson, E.
        2008, A\&A, 491, 379
\bibitem[{Girardi et al.}(2016)]{Girardi2016} Girardi, M., Boschin, W., Gastaldello, F., et al. 2016, MNRAS, 456, 2829
\bibitem[{Golovich et al.}(2017)]{Golovich2017} Golovich, N., van Weeren, R. J., Dawson, W. A., Jee, M. J., \&
        Wittman, D. 2017, ApJ, 838, 110
\bibitem[{Gupta et al.}(2018)]{Gupta2018} Gupta, A., Yuan, T., Torrey, P., et al. 2018, MNRAS, 477, L35
\bibitem[{Hayashi \& White}(2006)]{Hayashi2006} Hayashi, E., \& White, S. D. M. 2006, MNRAS, 370, L38
\bibitem[{Joshi et al.}(2020)]{Joshi2020} Joshi, G. D., Pillepich, A., Nelson, D., et al. 2020, MNRAS, 496, 2673
\bibitem[{Lage \& Farrar}(2014)]{Lage2014} Lage, C., \& Farrar, G. 2014, ApJ, 787, 144
\bibitem[{Louren\c{c}o et al.}(2020)]{Lourenco2020} Louren\c{c}o, A. C. C., Lopes, P. A. A., Lagan\'{a}, T. F., et al.
        2020, MNRAS, 498, 835
\bibitem[{{\L}okas}(2020)]{Lokas2020} {\L}okas, E. L. 2020, A\&A, 638, A133
\bibitem[{Marinacci et al.}(2018)]{Marinacci2018} Marinacci, F., Vogelsberger, M., Pakmor, R., et al. 2018,
        MNRAS, 480, 5113
\bibitem[{Markevitch et al.}(2002)]{Markevitch2002} Markevitch, M., Gonzalez, A. H., David, L., et al.
        2002, ApJ, 567, L27
\bibitem[{Markevitch et al.}(2003)]{Markevitch2003} Markevitch, M., Mazzotta, P., Vikhlinin, A., et al. 2003,
        ApJ, 586, L19
\bibitem[{Markevitch et al.}(2004)]{Markevitch2004} Markevitch, M., Gonzalez, A. H., Clowe, D., et al. 2004,
        ApJ, 606, 819
\bibitem[{Markevitch et al.}(2005)]{Markevitch2005} Markevitch, M., Govoni, F., Brunetti, G., \& Jerius, D.
        2005, ApJ, 627, 733
\bibitem[{Mastropietro \& Burkert}(2008)]{Mastropietro2008} Mastropietro, C., \& Burkert, A. 2008, MNRAS, 389, 967
\bibitem[{Mei et al.}(2007)]{Mei2007} Mei, S., Blakeslee, J. P., C\^{o}t\'{e}, P., et al. 2007, ApJ, 655, 144

\bibitem[{Naiman et al.}(2018)]{Naiman2018} Naiman, J. P., Pillepich, A., Springel, V., et al., 2018, MNRAS, 477, 1206
\bibitem[{Nelson et al.}(2018)]{Nelson2018} Nelson, D., Pillepich, A., Springel, V., et al. 2018, MNRAS, 475, 624
\bibitem[{Nelson et al.}(2019)]{Nelson2019} Nelson, D., Springel, V., Pillepich, A., et al. 2019,
        Computational Astrophysics and Cosmology, 6, 2
\bibitem[{Pillepich et al.}(2018)]{Pillepich2018} Pillepich, A., Nelson, D., Hernquist, L., et al. 2018,
        MNRAS, 475, 648
\bibitem[{Planck Collaboration}(2013)]{Planck2013} Planck Collaboration 2013, A\&A, 554, A140
\bibitem[{Rodr\'{i}guez-Gonz\'{a}lvez et al.}(2011)]{Rodriguez2011} Rodr\'{i}guez-Gonz\'{a}lvez, C., Olamaie, M.,
        Davies, M. L., et al. 2011, MNRAS, 414, 3751
\bibitem[{Ryu et al.}(2003)]{Ryu2003} Ryu, D., Kang, H., Hallman, E., \& Jones, T. W. 2003, ApJ, 593, 599
\bibitem[{Sales et al.}(2020)]{Sales2020} Sales, L. V., Navarro, J. F., Penafiel, L., et al. 2020, MNRAS, 494, 1848
\bibitem[{Schaal \& Springel}(2015)]{Schaal2015} Schaal, K., \& Springel, V. 2015, MNRAS, 446, 3992
\bibitem[{Schaal et al.}(2016)]{Schaal2016} Schaal, K., Springel, V., Pakmor, R., et al. 2016, MNRAS, 461, 4441
\bibitem[{Springel}(2010)]{Springel2010} Springel, V. 2010, MNRAS, 401, 791
\bibitem[{Springel \& Farrar}(2007)]{Springel2007} Springel, V., \& Farrar, G. R. 2007, MNRAS, 380, 911
\bibitem[{Springel et al.}(2001)]{Springel2001} Springel, V., White, S. D. M., Tormen, G., \& Kauffmann, G.
        2001, MNRAS, 328, 726
\bibitem[{Springel et al.}(2018)]{Springel2018} Springel, V., Pakmor, R., Pillepich, A., et al. 2018, MNRAS, 475, 676
\bibitem[{Tempel et al.}(2017)]{Tempel2017} Tempel, E., Tuvikene, T., Kipper, R., \& Libeskind, N. I. 2017,
        A\&A, 602, A100
\bibitem[{Thompson et al.}(2015)]{Thompson2015} Thompson, R., Dav\'{e}, R., Nagamine, K. 2015, MNRAS, 452, 3030
\bibitem[{Ubertosi et al.}(2023)]{Ubertosi2023} Ubertosi, F., Gitti, M., Brighenti, F., et al. 2023, ApJ, 944, 216
\bibitem[{Yun et al.}(2019)]{Yun2019} Yun, K., Pillepich, A., Zinger, E., et al. 2019, MNRAS, 483,
1042


\end{thebibliography}
\end{document}